\documentclass[camera,hyphens]{jpaper}
\usepackage{cite}
\usepackage{amsmath,amssymb,amsfonts}
\usepackage{graphicx}
\usepackage{fancyhdr}
\usepackage{url}
\usepackage{color}
\usepackage[dvipsnames]{xcolor}
\usepackage{tikz}
\usepackage[ddmmyyyy]{datetime} 
\usepackage{algorithm}
\usepackage{comment}
\usepackage[noend]{algpseudocode}
\usepackage{setspace}
\makeatletter
\algrenewcommand\ALG@beginalgorithmic{\footnotesize}
\makeatother

\algrenewcommand{\algorithmiccomment}[1]{\hfill$\triangleright$ \textit{#1}}
\algrenewcommand\algorithmicindent{1.0em}

\algnewcommand{\IIf}[1]{\State\algorithmicif\ #1\ \algorithmicthen}
\algnewcommand{\EndIIf}{\unskip\ \ }

\usetikzlibrary{positioning}

\definecolor{blue(pigment)}{rgb}{0.2, 0.2, 0.7}

\definecolor{dgreen}{rgb}{0.00, 0.75, 0.00}
\definecolor{ddgreen}{rgb}{0.00, 0.50, 0.00}
\definecolor{ddred}{rgb}{0.50, 0.00, 0.00}

\def\BibTeX{{\rm B\kern-.05em{\sc i\kern-.025em b}\kern-.08em
    T\kern-.1667em\lower.7ex\hbox{E}\kern-.125emX}}

\definecolor{MidnightBlue}{rgb}{0.1, 0.1, 0.44}
\newcommand{\versionnum}[0]{3.5~---~\today~@~\currenttime~CET} 
\newif\ifcameraready
\camerareadyfalse

\fancypagestyle{firstpage}{
  \fancyhf{}

  \fancyfoot[C]{\thepage}
}  

\definecolor{mGreen}{rgb}{0,0.6,0}
\definecolor{mGray}{rgb}{0.5,0.5,0.5}
\definecolor{mPurple}{rgb}{0.58,0,0.82}
\definecolor{backgroundColour}{rgb}{0.95,0.95,0.92}
\definecolor{gray97}{gray}{0.97}

\newcommand*\circled[1]{\tikz[baseline=(char.base)]{
            \node[shape=circle,fill,inner sep=1pt] (char) {\textcolor{white}{#1}};}}

\title{NATSA: A Near-Data Processing Accelerator \\ for Time Series Analysis} 

\newcommand{\affilUMA}[0]{\textsuperscript{\S}}
\newcommand{\affilETH}[0]{\textsuperscript{$\ddagger$}}
\newcommand{\affilNTUA}[0]{\textsuperscript{$\dagger$}}

\author{
{Ivan Fernandez\affilUMA}\qquad~~~%
{Ricardo Quislant\affilUMA}\qquad~~~%
{Christina Giannoula\affilNTUA}\qquad~~~%
\vspace{2pt}
{Mohammed Alser\affilETH}\\
{Juan Gómez-Luna\affilETH}\qquad~~~%
{Eladio Gutiérrez\affilUMA}\qquad~~~%
{Oscar Plata\affilUMA}\qquad~~~%
\vspace{6pt}
{Onur Mutlu\affilETH}\\%
\emph{{\affilUMA University of Malaga \qquad  \qquad \affilNTUA National Technical University of Athens \qquad \affilETH ETH Z{\"u}rich
}}%
}

\begin{document}
\bstctlcite{IEEEexample:BSTcontrol}

\maketitle

\fancyhead{}
\ifcameraready
 \thispagestyle{plain}
 \pagestyle{plain}
\else
 \fancyhead[C]{\textcolor{MidnightBlue}{\emph{ICCD 2020 Camera Ready Version \versionnum}}}
 \fancypagestyle{firststyle}
 {
   \fancyhead[C]{\textcolor{black}{\emph{To appear in the 38$^{th}$ IEEE International Conference on Computer Design (ICCD 2020)}}}
   \fancyfoot[C]{\thepage}
 }
 \thispagestyle{firststyle}
 \pagestyle{firststyle}
\fi

\begin{abstract}
Time series analysis is a key technique for extracting and predicting events in domains as diverse as epidemiology, genomics, neuroscience, environmental sciences, economics, and more. \emph{Matrix profile}, the state-of-the-art algorithm to perform time series analysis, computes the most similar subsequence for a given query subsequence within a sliced time series. \emph{Matrix profile} has low arithmetic intensity, but it typically operates on large amounts of time series data. In current computing systems, this data needs to be moved between the off-chip memory units and the on-chip computation units for performing \emph{matrix profile}. This causes a major performance bottleneck as data movement is extremely costly in terms of both execution time and energy.

In this work, we present \emph{NATSA}, the \emph{first} Near-Data Processing accelerator for time series analysis. The key idea is to exploit modern \mbox{3D-stacked} High Bandwidth Memory (HBM) to enable efficient and fast specialized \emph{matrix profile} computation near memory, where time series data resides. NATSA provides three key benefits: 1) quickly computing the \emph{matrix profile} for a wide range of applications by building specialized energy-efficient floating-point arithmetic processing units close to HBM, 2) improving the energy efficiency and execution time by reducing the need for data movement over slow and energy-hungry buses between the computation units and the memory units, and 3) analyzing time series data at scale by exploiting low-latency, high-bandwidth, and energy-efficient memory access provided by HBM. Our experimental evaluation shows that NATSA improves performance by up to 14.2$\times$ (9.9$\times$ on average) and reduces energy by up to 27.2$\times$ (19.4$\times$ on average), over the state-of-the-art multi-core implementation. NATSA also improves performance by 6.3$\times$ and reduces energy by 10.2$\times$ over a general-purpose NDP platform with 64 in-order cores.
\end{abstract}

\section{Introduction}\label{sec:intro}
A time series is a chronologically ordered set of samples of a real-valued variable that can contain millions of observations. Time series analysis is used to analyze information in a wide variety of domains~\cite{SS17}: epidemiology, genomics, neuroscience, medicine, environmental sciences, economics, and more. Time series analysis includes finding similarities (\emph{motifs}~\cite{CKL03}) and anomalies (\emph{discords}~\cite{KLL+06}) between every two subsequences (i.e., slices of consecutive data points) of the time series~\cite{YZU+18,TL17}. There are two major approaches for motif and discord discovery: approximate and exact algorithms~\cite{M14}. Approximate algorithms~\cite{CKL03} are faster than exact algorithms, but they can provide inaccurate results or limited discord detection, which cannot be tolerated by many applications (e.g., vehicle safety systems~\cite{SBP+12}). 
Unlike approximate algorithms, exact algorithms~\cite{MKZ+09} do not yield false positives or discordant dismissals, but can be very time-consuming on large time series data. 
Thus, \emph{anytime} versions (aka interruptible algorithms) of exact algorithms are proposed to provide approximate solutions quickly~\cite{MPROFILEI,MPROFILEXI} and can return a valid result even if the user stops their execution early.
\vspace{-1mm}

\begin{figure}[h!]
  \centering
  \includegraphics[width=0.9\linewidth]{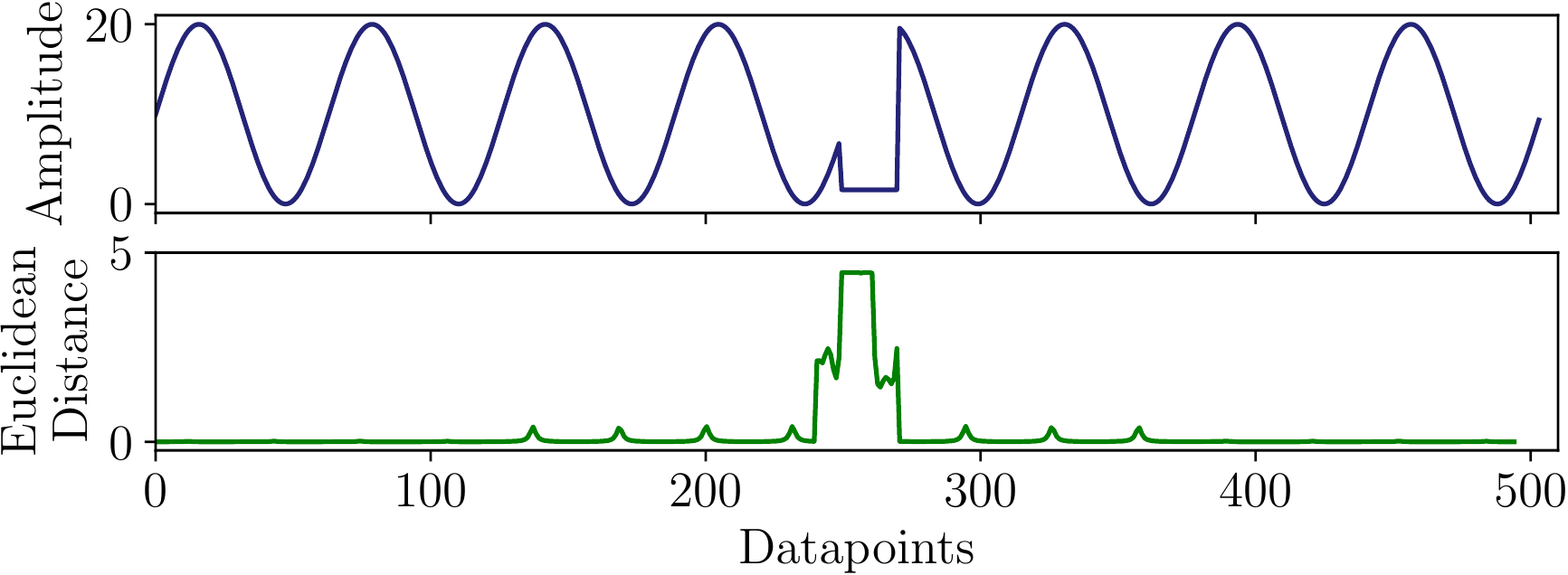}
  \caption{A time series (upper figure) including anomalies and its \emph{matrix profile} output (lower figure). Anomalies appear as higher Euclidean distance values in the profile.}
  \label{fig:mp_example}
\end{figure}

The state-of-the-art exact \emph{anytime} method for motif and discord discovery is {\it matrix profile}~\cite{MPROFILEI}, which is based on Euclidean distances and floating-point arithmetic. Fig.~\ref{fig:mp_example} depicts a naive example of anomaly detection using \emph{matrix profile}, where the sinusoidal signal has an anomaly between values 250 and 270. The \emph{matrix profile} output of this time series shows low values for the periodic subsequences of it as they are very similar to the other subsequences, and higher values for the anomalies and their neighboring subsequences.

We evaluate a recent CPU implementation of the \emph{matrix profile} algorithm~\cite{MPROFILEXI} on a real multi-core machine (Intel Xeon Phi KNL~\cite{sodani2015knights}) and observe that its performance is heavily bottlenecked by data movement. In other words, the amount of computation per data access is not enough to hide the memory latency and thus time series analysis is  memory-bound. This overhead caused by data movement limits the potential benefits of acceleration efforts that do not alleviate data the movement bottleneck in current time series applications. 
\vspace{-1mm}

Several CPU and GPU implementations of \emph{matrix profile} have been proposed in the literature~\cite{MPROFILEI,MPROFILEII,MPROFILEXI, MPROFILEXIV}. However, these acceleration efforts still require transferring the time series data from the main memory to the CPU/GPU cores, leading to the data movement bottleneck. 
Near-Data Processing (NDP)~\cite{mutlu2019processing, ghose2018enabling, ghose2019processing,Hsieh2016accelerating,wong2012metal,aga2017compute,asghari2016chameleon,chi2016prime,hashemi2016continuous,KKC+16,loh2013processing,seshadri2019dram,ahn2015pim,seshadri2017simple, GAK15, DDM+17, AHY+15, GPY+17, LMd+18, Boroumand2018google, boroumand2019conda, boroumand2017lazypim, Hsieh2016TOM,Kim2018Grim,cali2020genasm,hashemi2016accelerating,seshadri2015fast,li2016pinatubo,singh2019napel,pattnaik2016scheduling,seshadri2017ambit,singh2020nero,hashemi2016accelerating,seshadri2013rowclone,qiao2018atomlayer,chang2016low,kim2018dram,kim2019d,kim2018dram,kim2019d} is a promising approach to alleviate data movement by placing processing units close to memory. As a result, NDP solutions have the potential to improve system performance and energy efficiency when they are carefully designed with low-cost and low-overhead near data processing cores for memory-bound applications~\cite{GAK15, Boroumand2018google, AHY+15, ahn2015pim, mutlu2019processing, ghose2019processing, ghose2018enabling,Hsieh2016accelerating,GPY+17,LMd+18,Kim2018Grim,boroumand2019conda,PJZ+14,mutlu2003runahead,1261383}.

Our \textbf{goal} in this work is to enable high-performance and energy-efficient time series analysis for a wide range of applications, by minimizing the overheads of data movement. This can enable efficient time series analysis on large-scale systems as well as embedded and mobile devices, where power consumption is a critical constraint (e.g., heart beat analysis on a mobile medical device to predict a heart attack~\cite{LWT+19}).
\textbf{To this end}, we propose \emph{NATSA}, the \emph{first} \underline{N}ear-Data Processing \underline{A}ccelerator for \underline{T}ime \underline{S}eries \underline{A}nalysis. The key idea of NATSA is to exploit modern \mbox{3D-stacked} High Bandwidth Memory (HBM)~\cite{HBM,Lee2016Simultaneous} along with specialized custom processing units in the logic layer of HBM, to enable energy-efficient and fast \emph{matrix profile} computation near memory, where time series data resides. NATSA supports a wide range of time series applications thanks to \emph{matrix profile}'s generality and flexibility. 

Our evaluation shows that NATSA provides up to 14.2$\times$ (9.9$\times$ on average) higher performance and up to 27.2$\times$ (19.4$\times$ on average) lower energy consumption compared to a state-of-the-art multi-core system. NATSA consumes 11.0$\times$ and 4.1$\times$ less energy over optimized implementations of \emph{matrix profile} on an Intel Xeon Phi KNL~\cite{FVG+19} and NVIDIA GTX 1050 GPU~\cite{MPROFILEII}, respectively. NATSA has 9.6$\times$ and 1.8$\times$ smaller area than these two accelerators, at equivalent performance points. NATSA outperforms a general-purpose NDP platform 
by 6.3$\times$ while consuming 10.2$\times$ less energy. 

This work makes the following \textit{contributions}:
\begin{itemize}
    \item We propose \emph{NATSA}, the first near-data processing accelerator for accelerating time series analysis using modern 3D-stacked High Bandwidth Memory (HBM).
    \item We propose a new workload partitioning scheme that preserves the \emph{anytime} property of the algorithm, while providing load balancing among near-data processing units.
    \item We perform a detailed analysis of NATSA in terms of both performance and energy consumption. We compare different versions of NATSA (DDR4~\cite{DDR4} and HBM~\cite{HBM}) with four different architectures (8-core CPU, 64-core CPU, GPUs and NDP-CPU) and find that NATSA provides the highest performance and lowest energy consumption.
\end{itemize}

\section{Background}
\subsection{Time Series Analysis: The \emph{matrix profile}} 
\label{sec:mprofile}
A {\em time series} $T$ is a sequence of $n$ data points $t_{i}$, where $1\leq i\leq n$, collected over time. A subsequence of $T$, also called a \textit{window}, is denoted by $T_{i,m}$, where $i$ is the index of the first data point, and $m$ is the number of samples in the subsequence, with $1\leq i$, and $m\leq n - i$.

The state-of-the-art exact \emph{anytime} method for time series analysis is {\it matrix profile}~\cite{MPROFILEI}. When analyzing a time series, the \emph{profile} is maintained as another time series that represents the most similar neighbor for a particular subsequence of the original time series. The similarity between two subsequences $T_{i,m}$ and $T_{j,m}$ can be calculated using the \emph{z-normalized Euclidean distance}, which is defined as follows.
\vspace{-2mm}
\begin{equation}
    d_{i,j} =  \sqrt{2m \left( 1- \frac{Q_{i,j}-m\mu_i\mu_j}{m\sigma_i\sigma_j} \right)}
    \label{eq:dist}
    \vspace{-2mm}
\end{equation}

\noindent where $Q_{i,j}$ is the dot product of $T_{i,m}$ and $T_{j,m}$; $\mu_x$ and $\sigma_x$ are the mean and the standard deviation of the points in $T_{x,m}$, respectively. These statistics are computed in $O(n)$ time~\cite{MILLIONTRILLIONS}.

Using the distance in Eq.~\ref{eq:dist}, the \emph{matrix profile} algorithm solves the similarity search problem in three steps. First, it builds a symmetric $(n-m+1) \times (n-m+1)$ matrix $D$, called \textit{distance matrix}. Each cell in $D$, $d_{i,j}$, stores the distance between two subsequences, $T_{i,m}$ and $T_{j,m}$. Second, it creates an array $P$ of size $n-m+1$, called \emph{profile}. Each cell $P_i$ in $P$ keeps the minimum distance recorded in the $i^{th}$ row of $D$. Third, it allocates an array $I$ that is of the same size as $P$, called \emph{profile index}, such that $I_{i} = j$ if $P_{i} = d_{i,j}$. This way, $P$ contains the minimum distances between subsequences, while $I$ is the vector of ``pointers'' to the location of these subsequences within the time series.

\begin{figure}[h!]
    \centering
    \includegraphics[width=0.9\linewidth]{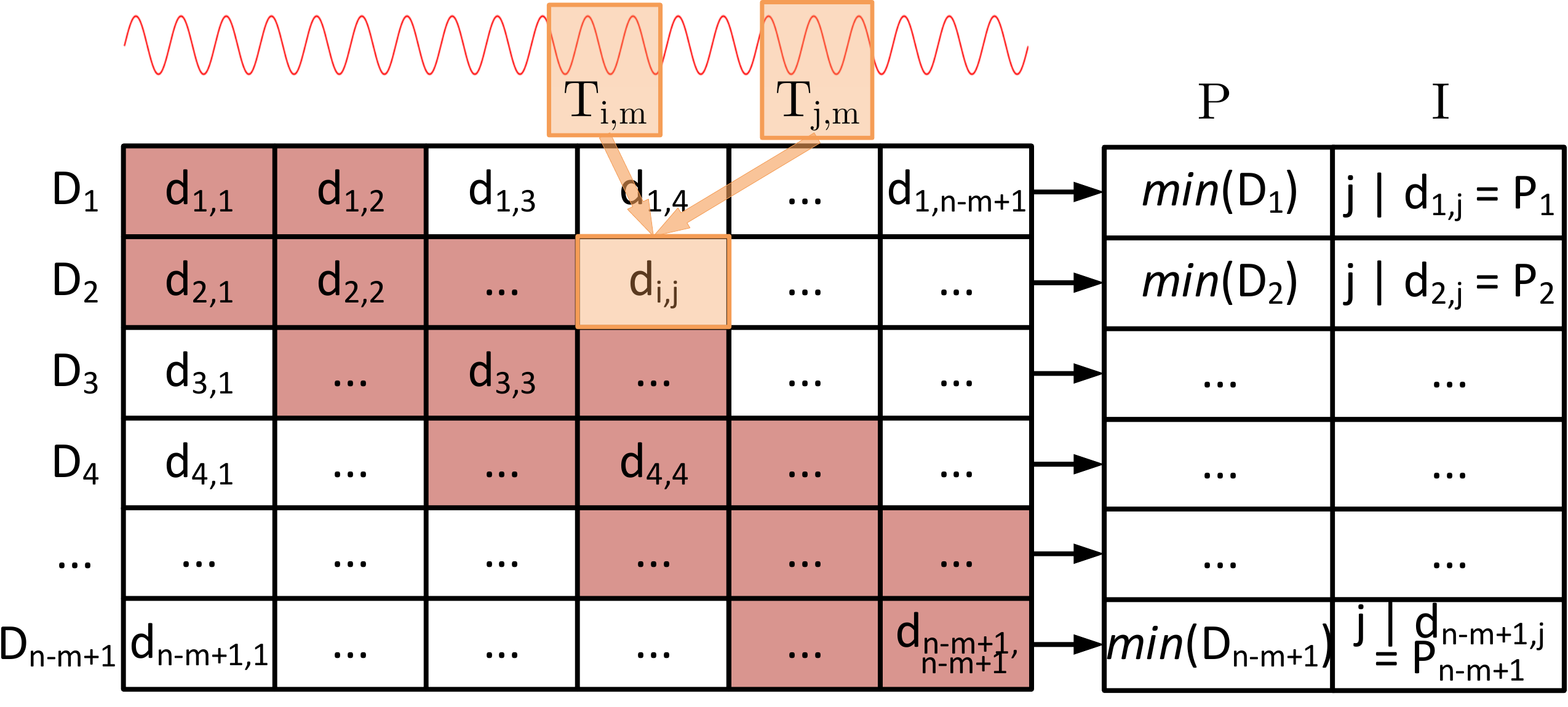}
    \caption{Example of distance matrix (D), profile (P), and profile index (I). $P_i$ holds the minimum distance calculated in row $D_i$, and $I_i$ holds the index $j$ of the subsequence that results in that distance. The cells in the exclusion zone are coloured red.}
    \label{fig:dist_example}
    \vspace{-2mm}
\end{figure}

Fig.~\ref{fig:dist_example} depicts an example of the distance matrix $D$, the profile $P$, and the profile index $I$. The neighboring subsequences of $T_{i,m}$ are highly similar to it (i.e., \mbox{$d_{i,i+1}\approx0$}) due to overlapping between them. The algorithm excludes these subsequences from the computation to avoid false positives, by defining an exclusion zone for each subsequence. It follows the approach in~\cite{MPROFILEXI}, where the exclusion zone of $T_{i,m}$ is $T_{i,\frac{m}{4}}$ (i.e., ends at $t_{i+\frac{m}{4}}$ of the time series). 

\subsection{The SCRIMP Implementation} \label{sec:SCRIMP}
The state-of-the-art CPU-based implementation of the \emph{matrix profile} algorithm is SCRIMP~\cite{MPROFILEXI}. We use an optimized version of SCRIMP~\cite{FVG+19} as baseline for our work, since it has the best convergence properties and takes advantage of multithreading and vectorization. The key mechanism behind optimized SCRIMP is that the dot product in Eq.~\ref{eq:dist} can be calculated incrementally in the diagonals of $D$ as follows:
\vspace{-2mm}
\begin{equation}
Q_{i,j} = Q_{i-1,j-1} - t_{i-1}t_{j-1} + t_{i+m-1}t_{j+m-1}
\label{eq:dist2}
\vspace{-2mm}
\end{equation}
According to Eq.~\ref{eq:dist2}, except for the first dot product, the remaining cells of a diagonal can be calculated using the values from the immediate upper left cells. This fact significantly reduces the number of multiplications and additions needed.

Algorithm~\ref{alg:scrimpUnroll}, optimized SCRIMP~\cite{FVG+19}, exploits both thread-level parallelism and vectorization. First, it precalculates the means and standard deviations of every subsequence of the time series (line~\ref{alg1:pre}), and initializes the profile vector (lines~\ref{alg1:Pinit_1}-\ref{alg1:Pinit}). Second, it computes the diagonals (see Fig.~\ref{fig:dist_example}) using the loop in line~\ref{alg1:for1}. The variable {\tt nDiag} is the number of diagonals of $D$ assigned to each thread. These diagonals can be ordered in the {\tt diag} vector (line~\ref{alg1:diag}) a) \emph{randomly}, enabling the \emph{anytime} property of the algorithm, or b) \emph{sequentially}, discarding the \emph{anytime} property but allowing for optimizations~\cite{MPROFILEXI} (e.g., exploiting data locality of consecutive diagonals). 

\begin{algorithm}[h!]

    \caption{Optimized SCRIMP~\cite{FVG+19}} \label{alg:scrimpUnroll}
    \begin{algorithmic}[1]
        \State $\mu, \sigma \leftarrow precalculateMeansDevs(T, m);$ \label{alg1:pre}
        \State $vectFact \leftarrow $ \textsc{vector\_width}$/sizeof(datatype);$
        \For {$i \leftarrow 0$ to $size(P)-1$} 
        \label{alg1:Pinit_1}
        \State $P_i \leftarrow  \infty;$
        \EndFor
        \label{alg1:Pinit}
        \For { $idx \leftarrow tid*nDiag$ to $(tid+1)*nDiag-1$} \label{alg1:for1}
            \State $i \leftarrow 0; j \leftarrow diag_{idx};$ \label{alg1:diag}
            \State $q \leftarrow dotProduct(T_{i,m},T_{j,m});$ \label{alg1:dotP} \Comment{Vectorized loop}
            \State $d \leftarrow dist(m,q,\mu_i,\sigma_i,\mu_j,\sigma_j);$ \label{alg1:dist}
            \IIf {$d < P_i$} \label{alg1:min} $P_i \leftarrow d; I_i \leftarrow j;$  \EndIIf
            \IIf {$d < P_j$}  $P_j \leftarrow d; I_j \leftarrow i;$ \EndIIf  \label{alg1:nim}
            \State $i \leftarrow i + 1;$
            \For {$j \leftarrow diag_{idx} + 1$ to $size(P)$}
                \For {$k \leftarrow 0$ to $vectFact - 1$} \Comment{Vectorized loop} \label{alg1:add}
                    \State $qs_{k} \leftarrow t_{i+m-1+k}t_{j+m-1+k} - t_{i-1+k}t_{j-1+k};$
                \EndFor \label{alg1:dda}
                \State $qs_{0} \leftarrow qs_{0} + q;$ \label{alg1:up}
                \For {$k \leftarrow 1$ to $vectFact - 1$} \label{alg1:seqUp}
                    \State $qs_{k} \leftarrow qs_{k} + qs_{k-1};$
                \EndFor \label{alg1:upSeq}
                \State $q \leftarrow qs_{vectFact - 1};$ \label{alg1:nextQ}
                \For {$k \leftarrow 0$ to $vectFact - 1$} \Comment{Vectorized loop} \label{alg1:distP}
                    \State $ds_{k} \leftarrow dist(m,qs_{k},\mu_{i+k},\sigma_{i+k},\mu_{j+k},\sigma_{j+k});$
                    \IIf {$ds_{k} < P_{i+k}$}
                         $P_{i+k} \leftarrow ds_{k}; I_{i+k} \leftarrow j+k;$
                    \EndIIf
                    \IIf {$ds_{k} < P_{j+k}$}
                         $P_{j+k} \leftarrow ds_{k}; I_{j+k} \leftarrow i+k;$  
                    \EndIIf
                \EndFor \label{alg1:Pdist}
                \State $i \leftarrow i + vectFact;$ 
            \EndFor
        \EndFor
    \end{algorithmic}
\end{algorithm}

Note that only $P$ and $I$ are allocated in memory, since storing $D$ can lead to large memory consumption for large series due to the $n^{2}$ memory footprint (i.e., the values of $D$ are calculated on the fly, updating $P$ and $I$ when needed). For each diagonal, the algorithm first computes the dot product of the first pair of subsequences in line~\ref{alg1:dotP} using the \textit{dotProduct} function, which is vectorized. Second, it calculates the distance according to Eq.~\ref{eq:dist} (line~\ref{alg1:dist}). Third, it checks and replaces the corresponding profile element with the new distance provided that the calculated one is smaller (lines~\ref{alg1:min}-\ref{alg1:nim}).

The algorithm addresses the imposed data dependency due to the dot product update between the elements in the diagonal with the following steps: 1) it pre-computes the add terms in Eq.~\ref{eq:dist2} in batches of size $vectFactor$ in a vectorized manner (lines~\ref{alg1:add}-\ref{alg1:dda}); 2) it adds the previous dot product to the first new one (line~\ref{alg1:up}); 3) it sequentially updates the remaining dot products in the batch (lines~\ref{alg1:seqUp}-\ref{alg1:upSeq}) saving the last one for the next iteration of the diagonal (line~\ref{alg1:nextQ}); 4) it computes the distance as well as the profile update in a vectorized way (lines~\ref{alg1:distP}-\ref{alg1:Pdist}). As a result, all loops are fully vectorized except the one in lines~\ref{alg1:seqUp}-\ref{alg1:upSeq}.

\subsection{NDP and 3D-Stacked Memory}
Near-Data Processing (NDP)~\cite{mutlu2019processing, ghose2018enabling, ghose2019processing,Hsieh2016accelerating,wong2012metal,aga2017compute,asghari2016chameleon,chi2016prime,hashemi2016continuous,KKC+16,loh2013processing,seshadri2019dram,ahn2015pim,seshadri2017simple, GAK15, DDM+17, AHY+15, GPY+17, LMd+18, Boroumand2018google, boroumand2019conda, boroumand2017lazypim, Hsieh2016TOM,Kim2018Grim,cali2020genasm,hashemi2016accelerating,seshadri2015fast,li2016pinatubo,singh2019napel,pattnaik2016scheduling,seshadri2017ambit,singh2020nero,hashemi2016accelerating,seshadri2013rowclone,qiao2018atomlayer,chang2016low,kim2018dram,kim2019d} is a promising paradigm to reduce the data movement between CPUs and memory by placing simple general-purpose processors~\mbox{\cite{Hsieh2016TOM,AHY+15,Boroumand2018google}} or application-specific accelerators~\cite{Hsieh2016accelerating,zhang2014top,Kim2018Grim, cali2020genasm,Boroumand2018google,ahn2015pim} in or close to the logic layer of 3D-stacked memory. Generally, NDP can provide performance benefits for memory-bound applications when they exhibit one or more of the following major properties: 1) requiring higher memory bandwidth than available in the system, 2) being sensitive to memory access latency~\cite{1261383}, or 3) performing irregular memory accesses, such that they cannot effectively benefit from cache hierarchy of conventional CPU architectures.
.

Recent advances in die-stacking technologies have enabled the integration of multiple layers of DRAM arrays in a single package. A 3D-stacked memory consists of several memory dies, one on top of each other, connected using Through-Silicon Vias (TSV)~\cite{HBM,Lee2016Simultaneous}. NDP locates low-power processing units inside the logic layer of 3D-stacked memory, to harness the significantly higher bandwidth and the lower latency provided while consuming less energy. The most prominent 3D-stacked memory technologies are High Bandwidth Memory (HBM)~\cite{JCL+17} and Hybrid Memory Cube (HMC)~\cite{hadidi2017demystifying}, but there are several others~\cite{KYM15,Ghose2019demystifying}.

\section{Motivation}\label{sec:motivation}
NATSA is motivated by two key observations:
First, time series motif and discord discovery are two of the most important analysis primitives for a wide variety of applications. Besides the applications mentioned in Section~\ref{sec:intro}, we can find these primitives applied to bioinformatics~\cite{B04,Alser2017GateKeeper,alser2020accelerating}, speech processing~\cite{GNN+17}, robotics~\cite{RGS+15}, weather prediction~\cite{MRB+11}, entomology~\cite{SDW15}, geophysics~\cite{CAC+13}, finance~\cite{CCR19}, communication engineering~\cite{LCD04}, and electroencephalography~\cite{HAA+17}.

\begin{figure}[h!]
    \centering
    \includegraphics[width=0.95\linewidth]{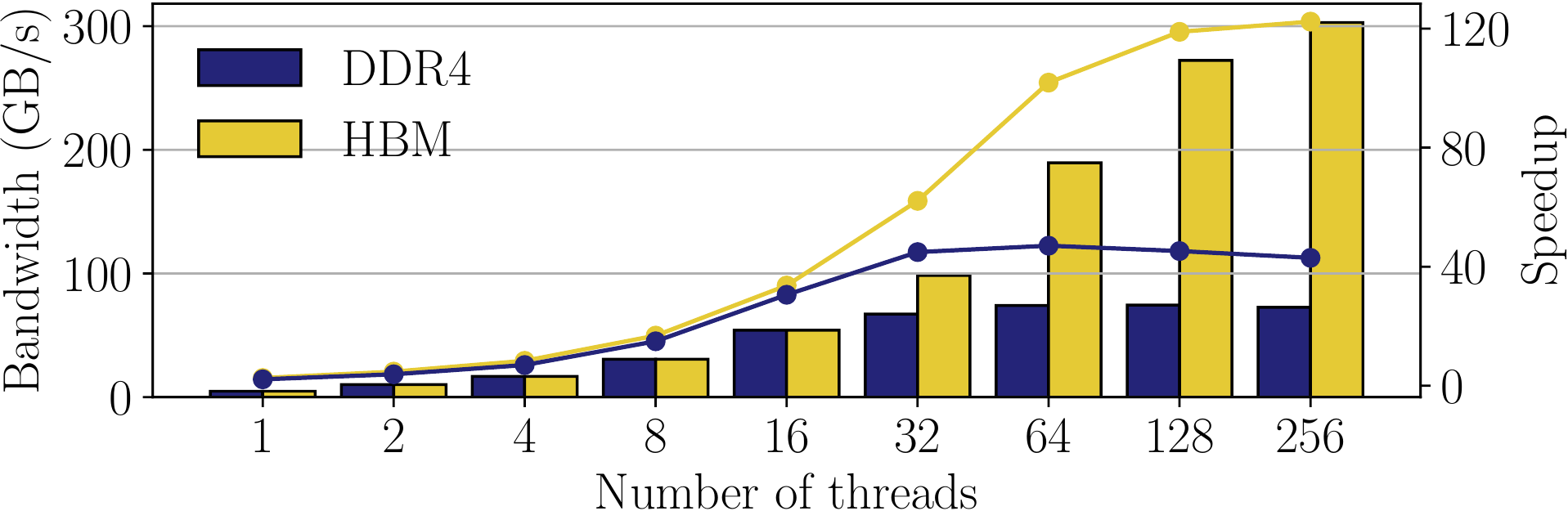}
    \caption{Memory bandwidth usage (bars) and normalized performance (lines) of a parallel and vectorized version of SCRIMP~\cite{FVG+19} running on an Intel Xeon Phi 7210.}
    \label{fig:HBMvsDDR}
\end{figure}

Second, memory is the main bottleneck in time series analysis. We characterize the performance of a state-of-the-art CPU-based multithreaded and vectorized implementation of  SCRIMP, developed in~\cite{FVG+19}. We run SCRIMP~\cite{FVG+19} on an Intel Xeon Phi 7210 processor, with 64 cores and 256 hardware threads, using two types of memory (DDR4 and HBM) available in this architecture. In Fig.~\ref{fig:HBMvsDDR}, we present the performance results normalized to 1 thread (lines) and utilized memory bandwidth (bars) of SCRIMP.
We observe that, when using DDR4, the performance of SCRIMP does not scale beyond 32 threads, whereas the higher memory bandwidth provided by HBM enables SCRIMP to scale up to 128 threads. This shows that SCRIMP's performance saturates on many-core architectures, because the achievable bandwidth saturates when the number of threads increases. 
To know the cause for this memory boundedness we perform the next experiment.

\begin{figure}[!h]
    \centering
    \includegraphics[width=0.9\linewidth]{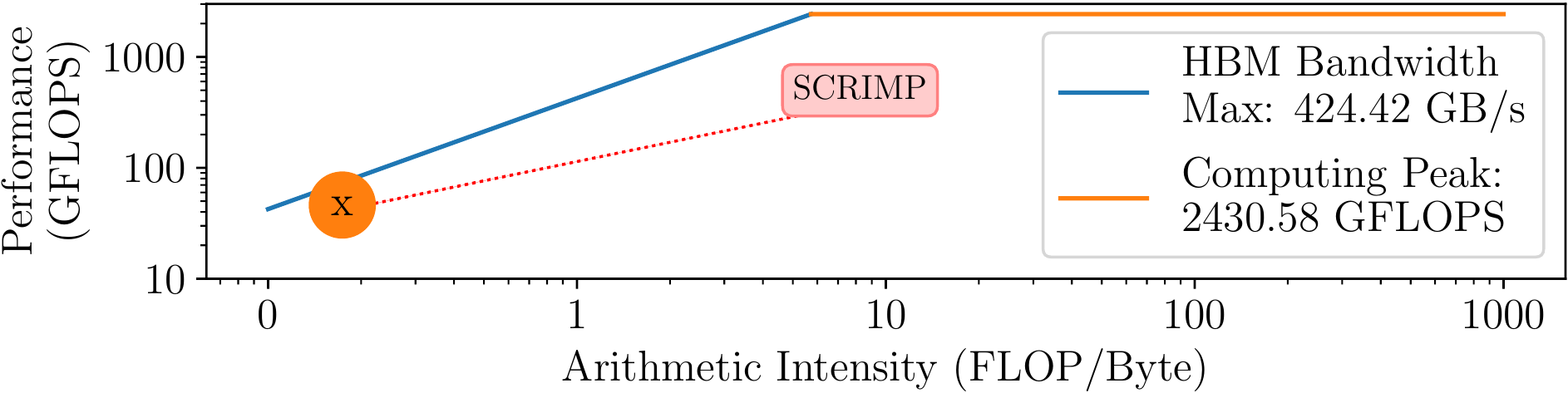}
    \caption{Roofline analysis of a parallel and vectorized version of SCRIMP~\cite{FVG+19} running on an Intel Xeon Phi 7210.}
    \label{fig:roofline}
\end{figure}

We perform the roofline analysis as we show in Fig.~\ref{fig:roofline}. 
We observe that the arithmetic intensity of SCRIMP is significantly low.
The confirms that the memory boundedness of SCRIMP is due to the low arithmetic intensity of the algorithm, which leads processing cores to be underutilized. Based on all these observations, we conclude that the performance of the state-of-the-art CPU-based implementation of the \emph{matrix profile}, SCRIMP~\cite{FVG+19}, is heavily bottlenecked by available memory bandwidth and data movement. Our goal is to reduce the data movement bottleneck of SCRIMP by building an NDP accelerator that matches the compute throughput of processing elements with the available memory bandwidth.

\section{NATSA Architecture}
\label{sec:natsa_overview}
Our \underline{N}ear-Data Processing \underline{A}ccelerator for \underline{T}ime \underline{S}eries \underline{A}nalysis, NATSA, is designed to 1) fully exploit the memory access parallelism and high memory bandwidth offered by HBM, and 2) employ the required amount of computing resources to provide a balanced solution. NATSA is built next to the HBM memory and exploits the full HBM bandwidth available. NATSA consists of multiple processing units (PUs) that efficiently compute the diagonals of \emph{matrix profile} in a parallel fashion. The PUs are designed to compute diagonals using a vectorized approach to process a batch of elements of a diagonal at the same time. Each PU includes energy-efficient floating-point units~\cite{GH10}, bitwise operators, and registers (See Table~\ref{tab:acc_param} in Sect.~\ref{sec:natsa_design}). Each PU communicates with the HBM memory via a controller connected to one of the 8 memory channels provided by HBM.

\subsection{NATSA Processing Units (PUs)}\label{seq:acc_pus}
\begin{figure}[h!]
  \centering
  \includegraphics[width=0.95\linewidth]{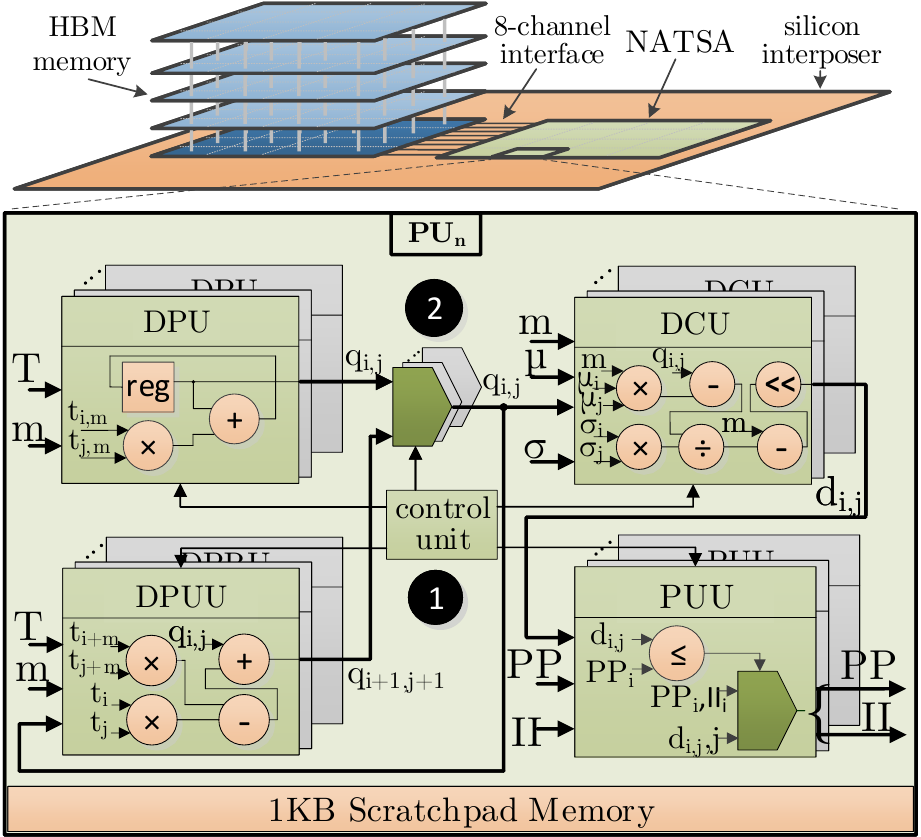}
   \caption{NATSA design and integration next to HBM memory. NATSA is connected directly to the HBM interface.}
  \label{fig:acc_datapath}
\end{figure}

Each NATSA PU consists of four hardware components: the \textit{Dot Product Unit} (DPU), the \textit{Distance Compute Unit} (DCU), the \textit{Profile Update Unit} (PUU), and the \textit{Dot Product Update Unit} (DPUU), as we show in Fig.~\ref{fig:acc_datapath}. We share the floating-point arithmetic operators (e.g., multipliers) among those hardware components to minimize idle cycles and enable reusability. The control unit (\circled{1} in Fig.~\ref{fig:acc_datapath}) is a state machine that orchestrates the execution flow of a PU. The multiplexers (\circled{2} in Fig.~\ref{fig:acc_datapath}) choose between the output of DPU and DPUU based on a signal from the control unit, so that the DCU can take advantage of Eq.~\ref{eq:dist2}, starting from the second element of the diagonal all the way down to the last. We replicate those hardware components to compute different elements of a diagonal in parallel, using the vectorized approach outlined in Section~\ref{sec:SCRIMP}. The diagonal assignment is pre-calculated in the host CPU, which sends the indices of the to-be-computed diagonals to each NATSA PU. Finally, each NATSA PU uses its own 1KB scratchpad memory to temporarily store fixed-size auxiliary data, such as the window size or configuration parameters.

The execution flow through the hardware components of a PU includes the following six steps: 
\begin{enumerate}
    \item \textbf{Dot product computation of the first element of the diagonal.} The DPU calculates the dot product between the first pair of subsequences of the diagonal ($T_{i,m}$ and $T_{j,m}$) by using the time series input, and the window size, \textsf{m}, which is used to signal the end of each subsequence. This hardware component
    vectorizes the operation and outputs the result, $q_{i,j}$, for the next step.

    \item \textbf{Euclidean distance computation of the first element of the diagonal}. The DCU computes the first Euclidean distance of each diagonal following Eq.~\ref{eq:dist}, using the dot product computed by the DPU $q_{i,j}$. 
    The values of $\mu$ and $\sigma$ are precomputed by the host CPU in negligible time ($O(n)$~\cite{MILLIONTRILLIONS}) with respect to the total execution time. This simplifies the design of the PU.
    
    \item \textbf{First profile update.} If the Euclidean distance calculated in the DCU, $d_{i,j}$, is lower than that stored in the profile for both subsequences, the PUU updates the profile vector and profile index vector, $PP$ and $II$.
   
    \item \textbf{Dot product update.} The dot product of the second and successive cells 
    in the diagonal is calculated from the previous cell. It is computed in the DPUU by subtracting the first product and adding the new one to $q_{i,j}$, as shown in Eq.~\ref{eq:dist2}. This hardware component is replicated to enable vectorization and is pipelined with the DCU and the PUU. 
    
    \item \textbf{Second and successive Euclidean distance computations.} The DCU computes again the Euclidean distance, but now it obtains $q_{i,j}$ from the DPUU. The DPUU hardware component is replicated for vectorization of the dot product update calculations.
    
    \item \textbf{Second and successive profile updates.} The PUU updates the profile vector and profile index vector, if needed. This hardware component is replicated to perform several updates at a time.
\end{enumerate}

\subsection{Workload Partitioning Scheme}\label{sec:diag_scheme}
Computing the diagonals of the distance matrix may lead to load imbalance among the PUs, because those diagonals have different lengths. To avoid this imbalance, we propose a static partition scheduling scheme which depends only on the size of the time series and the exclusion zone.

The way we tackle this problem is by assigning a set of pairs of diagonals to each NATSA PU such that the sum of their elements is equal to the number of cells of the main diagonal of the distance matrix minus the number of cells of the exclusion zone, $(n-m+1) - m/4$. 
 
\begin{figure}[h!]
    \centering
    \includegraphics[width=0.95\linewidth]{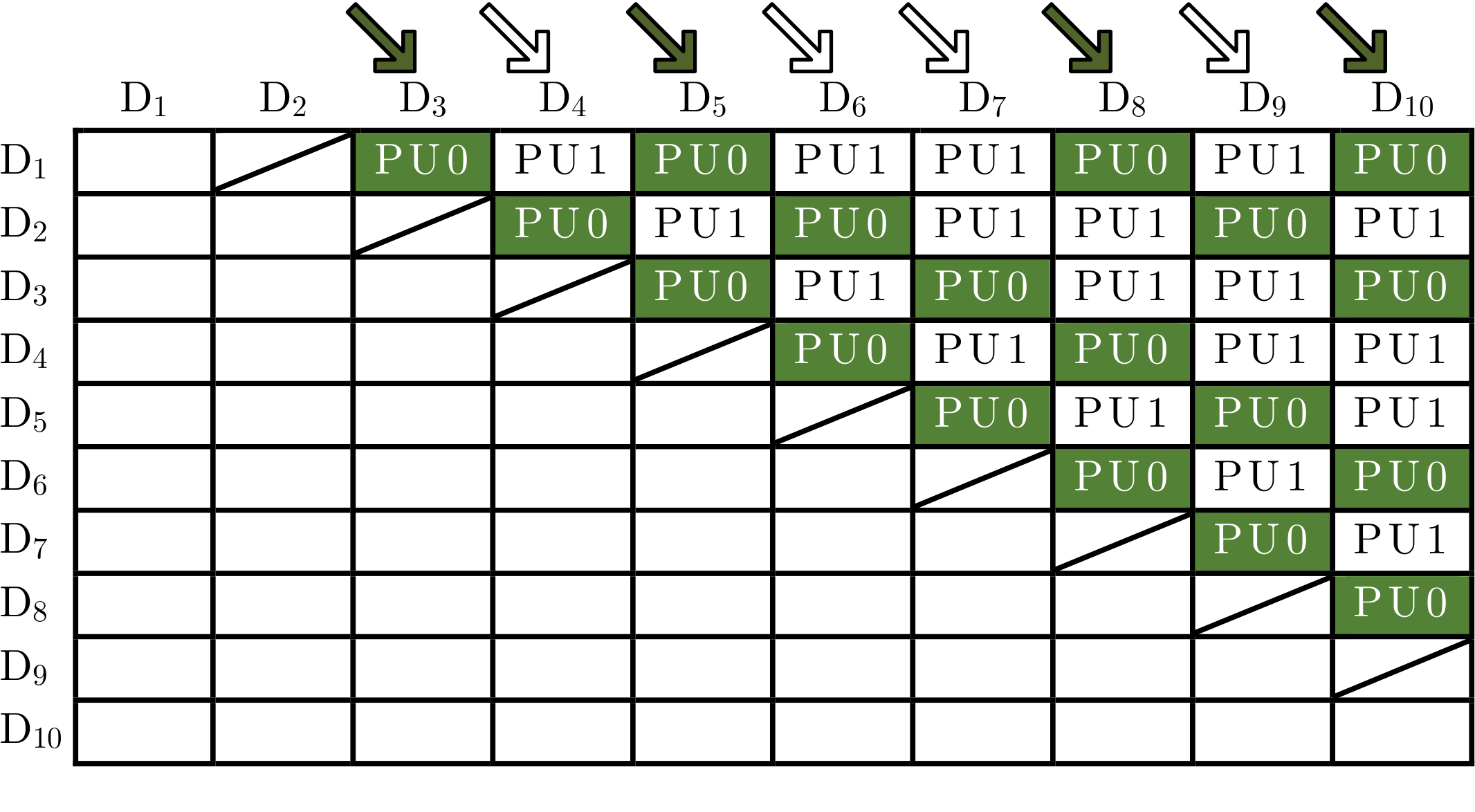}
    \caption{Example of the diagonal scheduling scheme for two processing units, denoted as \textit{PU0} (green) and \textit{PU1} (white). Arrows show direction of computation.}
    \label{fig:acc_sched}
\end{figure}

Fig.~\ref{fig:acc_sched} illustrates an example with two PUs, \textit{PU0} and \textit{PU1}, a distance matrix for a time series of $n=13$ cells, a window size of $m=4$, and an exclusion zone of 1 diagonal (crossed out rectangles). In this case, the number of elements that each pair of diagonals assigned to a PU should have is $(n-m+1) - m/4 = 10 - 1 = 9$. Comparing a subsequence with itself gives zero distance value. As a consequence, the algorithm treats the main diagonal as exclusion zone and avoids computing it. The first diagonal of non-zero values, which starts in column $D_{2}$ and is represented with crossed out rectangles, belongs to the exclusion zone (see Fig.~\ref{fig:dist_example}), so NATSA PUs also skip it.

Discarding the computation of the main diagonal and the diagonals in the exclusion zone, both PUs have to compute the diagonals from columns $D_{3}$ to $D_{10}$. To perform this efficiently and maintain the \emph{anytime} property of SCRIMP, in the first step, \textit{PU0} is assigned the first and last diagonal (9 elements in total), and \textit{PU1} is assigned the second and the penultimate diagonal (totalling 9 elements as well). In the second step, \textit{PU0} computes the third and the third-to-last diagonal, whereas \textit{PU1} computes the fourth and fifth diagonals.

Our proposed scheduling scheme can be used in two ways: 1) \emph{Randomly ordering} the indices of diagonals that each PU has to compute. Using this approach, we are able to preserve the \emph{anytime} property of the algorithm, since if the execution is interrupted, the user obtains a partial exploration of the whole time series (i.e., events from any point of the time series can be detected). 2) \emph{Sequentially ordering} the indices of diagonals that each PU has to compute. This approach violates the \emph{anytime} property (i.e., only events up to the interruption point can be detected), but allows for further optimizations (e.g., exploiting data locality between consecutive diagonals).

\textbf{Data mapping}. Each PU has access to its corresponding portion of the time series and statistic vectors, and works with replicated profile and profile index vectors. This approach simplifies the overall architecture, enabling the use of many PUs without having to synchronize between them. NATSA assigns multiple diagonals to each PU with the specific scheduling scheme described in this section. 

\subsection{Programming Interface} \label{sec:api}
In this section, we introduce the API to invoke NATSA from a host processor. While conventional loosely-coupled accelerators (e.g., GPUs or FPGAs) have their own memory, where data must be transferred to from the host's memory, NATSA is a tightly-integrated NDP accelerator, located between the host CPU and main memory. Thus, there is no need to transfer any data between the host memory and the accelerator memory, as loosely-coupled accelerators require. The user is responsible for 1) allocating the time series ($T$) and 2) providing the window length ($m$). NATSA will provide the user the profile vector ($P$) and profile index vector ($I$) in return. The size of the exclusion zone ($\frac{m}{4}$ by default) can be also passed as a parameter ($exc$).

Algorithm~\ref{alg:NATSA} outlines the \texttt{NATSA} API. First, \texttt{NATSA} function precalculates the statistics ($\mu,\sigma$) (line~2) in the host CPU and allocates the private vectors ($PP,II$) to NATSA's PUs (line~3).

\begin{algorithm}[h!]
    \caption{NATSA API} \label{alg:NATSA}
    \begin{algorithmic}[1]
    \Function{$P,I \leftarrow$ \texttt{NATSA}}{$T, m, exc, conf$}
        \State $\mu, \sigma \leftarrow precalculateMeanDev(T, m)$
        \State $PP,II \leftarrow allocatePrivateProfiles(T, m, exc)$
        \State $idx \leftarrow diagonalScheduling(T, m, exc)$
        \State \textsc{start\_accelerator}($T, m, exc, conf, idx, PP, II$)
        \State $P,I \leftarrow reduction(PP,II)$
    \EndFunction
    \end{algorithmic}
 
\end{algorithm}

Second, \texttt{NATSA} function implements the diagonal scheduling scheme presented in the previous section, setting the diagonals to be computed by each PU in $idx$ (line~4). Third, it initiates the accelerator (line~5), which starts the computation, and the host CPU waits for all the processing units to finish. Once the computation finishes, the host CPU performs the final reduction of the private vectors (line 6) and the user can find the results in the $P$ and $I$ vectors. The \texttt{conf} argument (line~1), besides holding configuration parameters for the accelerator, allows for future extensions, such as using other distance metrics (e.g., Pearson correlation~\cite{MPROFILEXIV}).

\section{Methodology}\label{sec:method}
We describe the simulation environment and the workload we use to evaluate the performance of NATSA.

\subsection{Simulation Environment}\label{sec:simEnv}
We simulate general-purpose cores using an in-house integration of \emph{ZSim}~\cite{SK13}, whose front-end is \emph{Pin}~\cite{LCM+05}, with \emph{Ramulator}~\cite{KYM15} \cite{RAMULATOR_GITHUB}. ZSim is a simulator which can model 1) general purpose cores (both in-order and out-of-order cores), and 2) the conventional cache hierarchy. Ramulator is a cycle-level and extensible DRAM simulator that provides a wide variety of memory models, including DDR4~\cite{DDR4} and HBM~\cite{HBM}. We use \emph{McPAT}~\cite{LAS+09} for power estimations.

For the NATSA accelerator, we use the \emph{gem5}~\cite{BBB+11} and \emph{Aladdin}~\cite{SRW+14} integration developed in~\cite{SXS+16}. Aladdin provides performance, area, and power estimations for a system-on-chip accelerator by requiring the equivalent C implementation of the accelerator design. Aladdin estimates the performance, power, and area of the accelerator within 0.9\%, 4.9\%, and 6.6\% compared to that provided by RTL flows, but over $100\times$ faster~\cite{SRW+14}. As Aladdin does not model the memory subsystem, we need to simulate it using gem5.

For a fair comparison, we evaluate our baseline platform (see the evaluated platforms below) in both ZSim and gem5 frameworks using the same workload (see Section~\ref{sec:mprofLoad}). We obtain up to 10\% simulated time reduction using ZSim with respect to gem5 (i.e., the baseline system performs slightly better with ZSim). As a consequence, the performance benefits of NATSA with respect to the baseline simulated using gem5, would be even higher. However, we choose ZSim since simulations of manycore systems with ZSim are orders of magnitude faster than gem5 simulations~\cite{SK13}, and this allows for the evaluation of general-purpose core platforms with large time series. For both general-purpose cores and accelerators, we obtain the power consumption of the memory system using the Micron Power Calculator~\cite{MICRONPOWERCALCS}, which we feed with the bandwidth usage from Ramulator and gem5, respectively. 

Using these simulation environments, we define several representative hardware platforms for the evaluation:
\begin{itemize}
    \item \textbf{DDR4-OoO (\textit{Baseline}):} A conventional DDR4-based system with eight four-wide out-of-order cores at 3.75GHz. Each core has 32KB private L1 instruction/data caches and a private 256KB L2 cache. The cores share an 8MB L3 cache. The main memory is a dual channel 16GB DDR4-2400 with 38.4GB/s of memory bandwidth.

    \item \textbf{DDR4-inOrder:} A conventional architecture using 64 in-order cores at 2.5GHz. Each core has only a single level of private 32KB instruction/data caches. The main memory is the same DDR4 as in the baseline system. We use this simple core-cache configuration to compare with the following NDP general-purpose-core system. 
    
    \item \textbf{HBM-OoO:} An NDP architecture with eight four-wide out-of-order cores at 3.75GHz. Each core has 32KB private L1 instruction/data caches and a private 256KB L2 cache. The main memory is a 4GB 3D-stacked HBM2 that provides a throughput of 256GB/s.

    \item \textbf{HBM-inOrder:} An NDP architecture with 64 in-order cores at 2.5GHz. Each core has a single level of private 32KB instruction/data caches. The main memory is a 4GB 3D-stacked HBM2 that provides a throughput of 256GB/s.

    \item \textbf{NATSA:} Our NDP accelerator with 48 PUs at 1GHz. Each PU has access to a private scratchpad memory of 1KB. The main memory is the same 4GB 3D-stacked HBM2 as in the \textit{HBM-OoO} and \textit{HBM-inOrder} platforms.
\end{itemize}

\subsection{Workload}
\label{sec:mprofLoad}
We use two real datasets and five synthetic datasets to evaluate the performance of NATSA against state-of-the-art architectures. The two real datasets are electrocardiogram (ECG) and seismology data obtained from~\cite{TDE+92} and \cite{YHK16}. We use these real datasets to 1) verify the correctness of the \emph{matrix profile} computed by NATSA (the same approach used in \cite{YHK16}) and 2) evaluate the effect of using single-precision versus double-precision (see Section~\ref{sec:accuracy_window}). We generate the five synthetic datasets of different representative lengths~\cite{MPROFILEXI} for performance evaluation using \mbox{MATLAB}, as shown in Table~\ref{tab:rand_sizes}. 

\begin{table} [h]
\caption{Synthetic time series for performance evaluation.}
\vspace{-1mm}
\label{tab:rand_sizes}
\resizebox{\linewidth}{!}{%
\begin{tabular}{l|ccccc}
\hline
        Time Series            &  \textbf{rand\_128K} & \textbf{rand\_256K} &  \textbf{rand\_512K} &  \textbf{rand\_1M} &  \textbf{rand\_2M} \\
        \hline
        Length (n) & 131072 &  262144 &  524288 &  1048576 & 2097152  \\
        \hline
    \end{tabular}}
\end{table}

\section{Evaluation}

In this section, we first evaluate NATSA's performance, comparing it to the general-purpose platforms (DDR4-OoO, DDR4-inOrder, HBM-OoO, and HBM-inOrder). Second, we compare NATSA to both simulated and real architectures (e.g., many-core CPUs and GPUs~\cite{MPROFILEII}) in terms of power consumption and area. Third, we present a design space exploration of NATSA. Fourth, we analyze the performance of general-purpose cores and their bottlenecks. Finally, we evaluate SCRIMP in terms of precision and sensitivity to subsequence lengths ($m$).

\subsection{Performance of NATSA}\label{sec:NATSAperf}

\begin{figure}[h!]
  \centering
  \includegraphics[width=0.95\linewidth]{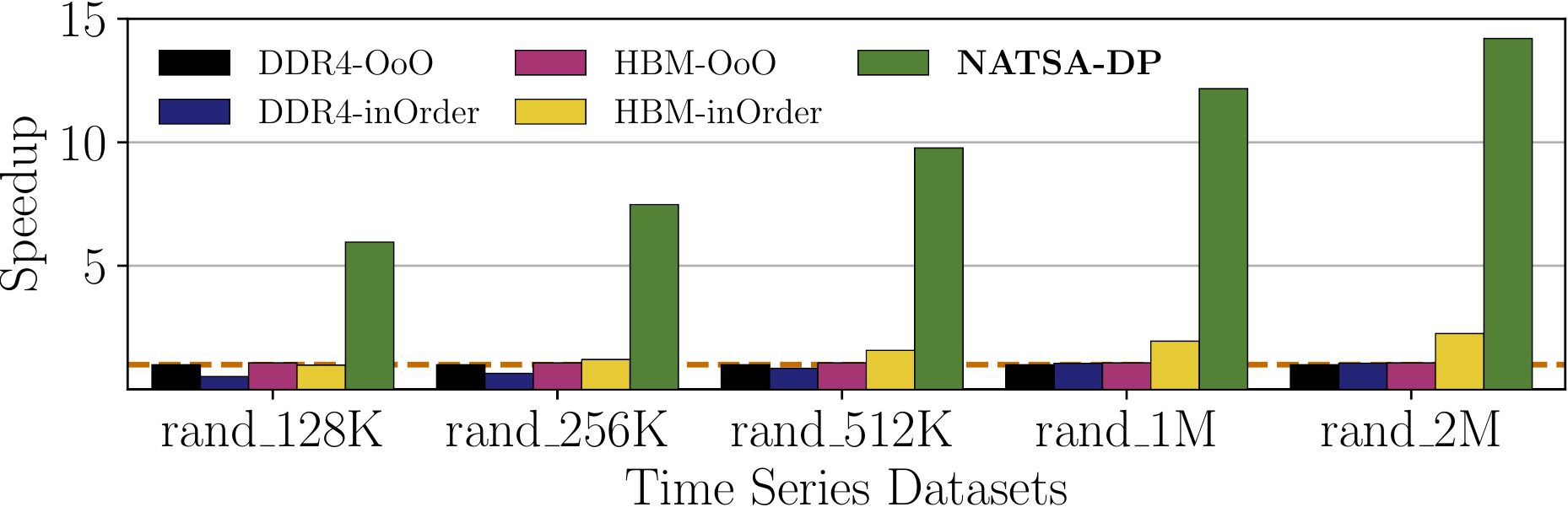}
  \caption{Speedup with respect to the baseline platform (DDR4-OoO) using double precision data.} 
  \label{fig:speedup_TISAN}
\end{figure}

We evaluate the performance of two NATSA designs using single-precision (SP) and double-precision (DP), respectively. We present normalized performance of NATSA-DP with respect to the baseline platform (DDR4-OoO) in Fig.~\ref{fig:speedup_TISAN}, using double-precision data. NATSA achieves significant performance improvements, up to 14.2$\times$ (9.9$\times$ on average) over the baseline system for large time series, and 6.3$\times$ over HBM-inOrder for all sizes. We observe that NATSA's speedup increases as the time series length becomes larger. This is because the arithmetic intensity decreases when the ratio of time series length ($n$) to window size ($m$) increases. Dot product update (Section~\ref{sec:SCRIMP}) causes the first dot product to take a significant part of the computation for shorter diagonals (lower $n$ to $m$ ratio). The cache hierarchy of the baseline system accelerates the first dot product. Conversely, a greater $n$ to $m$ ratio results in longer diagonals with the first dot product being less significant with respect to the total execution time, reducing the observed benefits of a cache hierarchy.

We evaluate the performance of the single-precision NATSA design.\footnote{We note that NATSA experiments are carried out with the gem5-Aladdin simulation framework, and the other platforms are evaluated with the ZSim-Ramulator framework (baseline system included). As mentioned in Section~\ref{sec:simEnv}, simulated times are slightly shorter for ZSim, so the actual gains of NATSA would likely be even greater what we report.} Table~\ref{tab:ooo_ddr4_exec} presents the average execution time for the analyzed datasets. NATSA-SP, which provides higher performance with similar area cost to NATSA-DP, outperforms NATSA-DP by up to 1.75$\times$, DDR4-OoO-DP by up to 24.9$\times$ and HBM-inOrder-DP by up to 11.1$\times$ for large time series.

\begin{table} [h!]
\caption{Execution time (in seconds) for single-precision and double-precision data.}
\vspace{-1mm}
\label{tab:ooo_ddr4_exec}
\resizebox{\linewidth}{!}{%
\begin{tabular}{l|rrrrr}
\hline
 \tikz[diag text/.style={inner sep=0pt, font=\footnotesize},
      shorten/.style={shorten <=#1,shorten >=#1}]{%
         \node[below left, diag text] (cfg) {\bf Config~~};
         \node[above right=0pt and 5pt, diag text] (n) {\bf ~~~Dataset};
         \draw[shorten=4pt, very thin] (cfg.center|-n.north west) -- (cfg.south east-|n.south east);}
                    & \bf rand\_128K & \bf rand\_256K & \bf rand\_512K & \bf rand\_1M & \bf rand\_2M \\
        \hline
        DDR4-OoO-DP &     14.72  &	77.55     &   414.55   & 2089.05     &  9810.30   \\
        DDR4-OoO-SP &     6.46   &	44.47     &   207.85   & 1106.36     &  5206.75   \\
        HBM-inOrder-DP &  14.95 &	64.20 & 262.33 &	1071.03 &	4347.38   \\
        HBM-inOrder-SP &    8.16 &	35.68 &	130.23 &	625.27 &	2466.69   \\
        NATSA-DP    &      2.47  &   10.37    &   42.45    &   171.72    &  690.65    \\
        \textbf{NATSA-SP}    &      \textbf{1.41}  &     \textbf{5.91}   &   \textbf{24.19}    &   \textbf{97.84}     & 	\textbf{393.45}    \\
        \hline
    \end{tabular}}
\end{table}
We conclude that NATSA provides the highest performance compared to modern general-purpose platforms.

\subsection{Power, Energy and Area Consumption}
\begin{figure}[h!]
	\centering
	\includegraphics[width=0.9\linewidth]{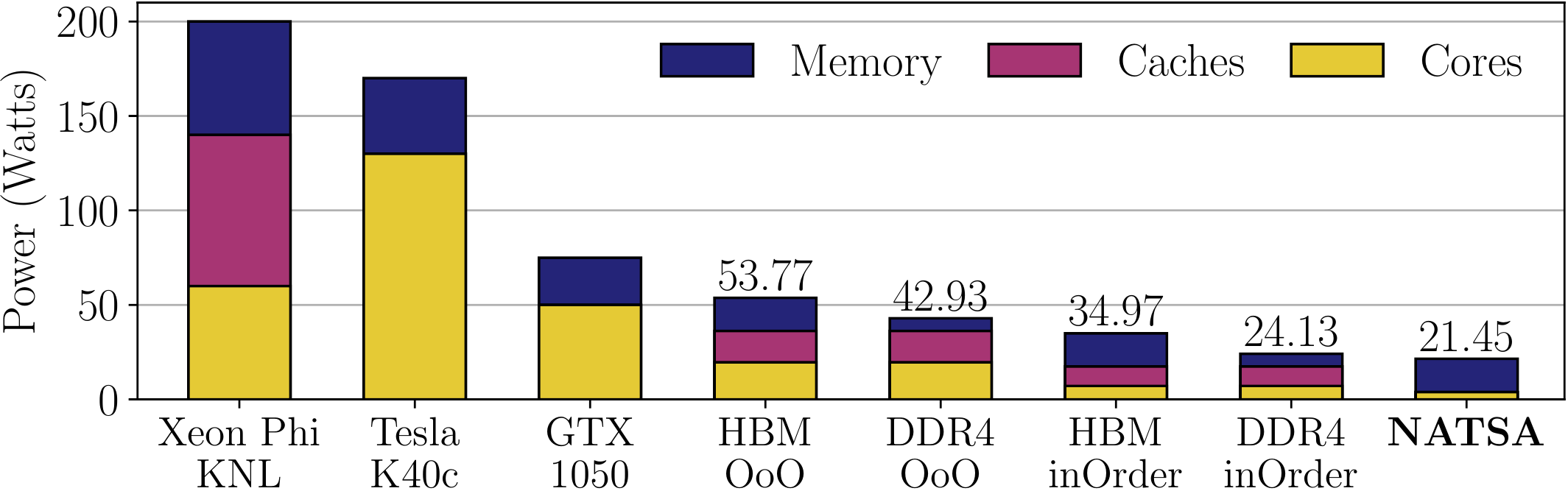}
	\caption{Dynamic power consumption for simulated and real hardware platforms.} 
	\label{fig:inst_energy}
\end{figure}

\begin{figure}[h!]
	\centering
	\includegraphics[width=0.9\linewidth]{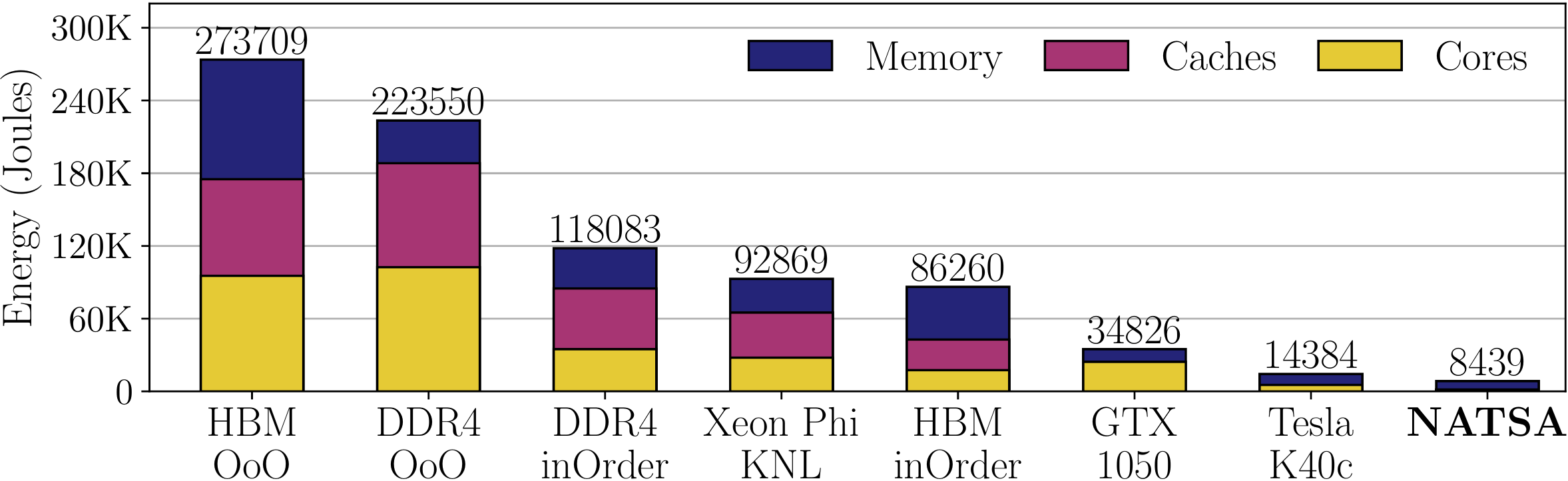}
	\caption{Energy consumption for simulated and real hardware platforms.}
	\label{fig:energy}
\end{figure}

\textbf{Power and Energy Consumption.} We compare the power and energy consumption of NATSA versus other existing hardware platforms in Figures \ref{fig:inst_energy} and \ref{fig:energy}. We use McPAT and Micron Power Calculators to evaluate energy consumption for the general-purpose platforms, getting the number of stalls and bandwidth usage from ZSim-Ramulator. For NATSA, we add Aladdin's energy estimations to the values obtained from the Micron Power Calculator. We also obtain energy measurements from real executions on GPUs using NVVP~\cite{NVVP} and on CPUs using PCM~\cite{PCM}, to compare NATSA with real platforms.

Fig.~\ref{fig:inst_energy} shows the dynamic power consumption of each simulated or real hardware platform. We observe that NATSA has the lowest power consumption, and most of its power is consumed by memory.

Fig.~\ref{fig:energy} shows the energy consumption of each simulated or real platform, for the computation of a time series of 524,288 elements (rand\_512K) using double-precision. 
To calculate the energy consumption, we compute the power-delay product with the measured instantaneous power consumption and the execution time. NATSA reduces energy consumption by 27.2$\times$ (19.4$\times$ on average) over the baseline platform (DDR4-OoO), and by 10.2$\times$ over an NDP architecture with general-purpose cores (HBM-inOrder). NATSA consumes 1.7$\times$, 4.1$\times$, and 11.0$\times$ less energy than an NVIDIA Tesla K40c GPU~\cite{nvidia2013k40}, NVIDIA GTX 1050 GPU~\cite{GTX1050}, and Intel Xeon Phi KNL~\cite{sodani2015knights}, respectively. We conclude that NATSA is the most energy-efficient evaluated platform for \emph{matrix profile}.

\begin{figure}[h!]
    \centering
    \includegraphics[width=0.9\linewidth]{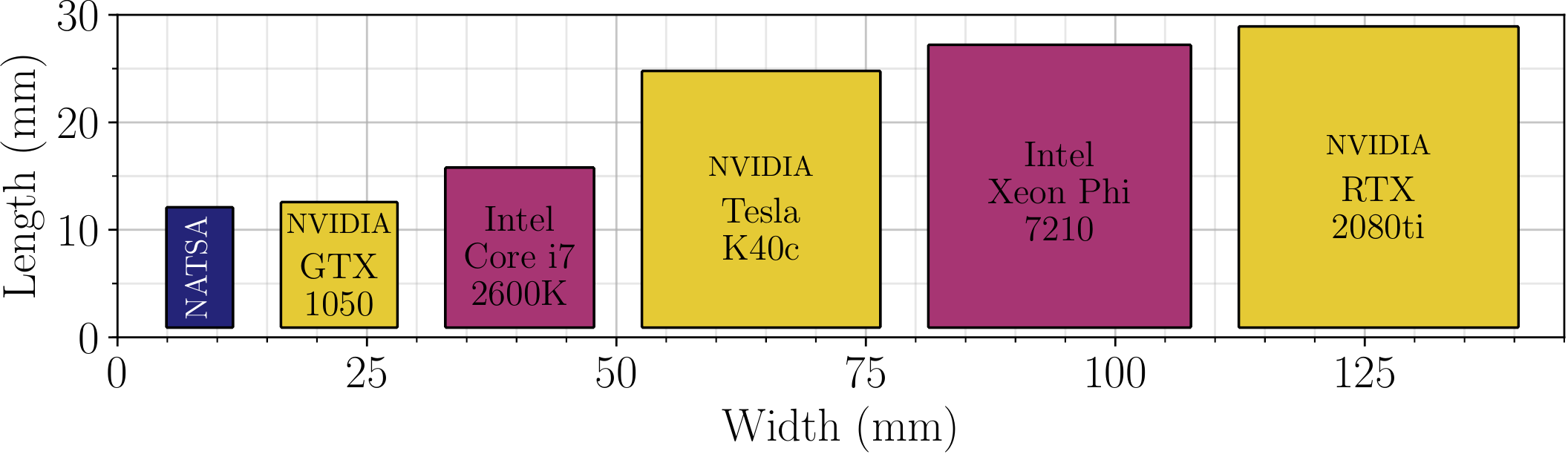}
    \caption{Area comparison of different hardware platforms.} 
    \vspace{-2mm}
    \label{fig:area}
\end{figure}

\textbf{Area}. We provide a scaled area comparison in Fig.~\ref{fig:area}. We observe that NATSA requires 9.6$\times$, 7.9$\times$, 3$\times$, and 1.8$\times$ less area than an Intel Xeon Phi KNL (14nm), NVIDIA Tesla K40c (28nm), Intel Core i7 (32nm), and NVIDIA GTX 1050 (14nm).

We conclude that NATSA (at 45nm technology node) is the platform that requires the least area, while using the largest technology node (i.e., 45nm) compared to other evaluated architectures. Using a more recent and smaller technology node (e.g., 15nm instead of 45nm) could additionally reduce NATSA's energy consumption by 4$\times$ and area by 3$\times$~\cite{salehi2015energy}.

\subsection{NATSA Design Space Exploration}
\label{sec:natsa_design}
We explore the key design choices of NATSA so that we deploy the exact number of PUs that saturate the memory bandwidth available, while minimizing the area and power consumption of the accelerator. We evaluate the use of HBM memory,\footnote{We also explore the use of DDR4 memory, where 8 PUs are enough to saturate the available memory bandwidth and the performance obtained is similar to the DDR4-inOrder platform (4\% difference).} where we find that 48 PUs make the accelerator balanced between memory bandwidth and compute parallelism, as 64 PUs result in a memory-bound accelerator, whereas 32 PUs a compute-bound one. Table~\ref{tab:acc_param} details the design parameters of NATSA for HBM. NATSA has 48 PUs which run at a frequency of 1GHz, fabricated at 45nm process. Implementations of NATSA with lower technology nodes would provide smaller area footprint and improved energy efficiency. Table~\ref{tab:acc_param} shows the components in a PU depending on the data precision: 1) double-precision (DP), and 2) single-precision (SP).

\begin{table} [h!]
    \caption{NATSA design components for 48 PUs.}\label{tab:acc_param}
    \vspace{-1mm}
    \resizebox{\linewidth}{!}{%
    \begin{tabular}{l||rrrr}
    \hline
        \bf Parameter/Component    & \bf PU-DP  & \bf NATSA-DP  & \bf PU-SP    & \bf NATSA-SP  \\
        \hline
         \hline
        Mem. bandwidth (GB/s)   & 5      & 240        & 5         & 240       \\
        Peak power (W)       & 0.1       & 4.8         & 0.08        & 3.84         \\
        Area ($mm^2$)    & 1.62        & 77.76          & 1.51          & 72.48          \\
        \hline
        FP Multipliers/Adders     & 16/14    &  768/672  &  64/36   &  3072/1728 \\
        Integer Adders  & 16    &  768  &  64   &  3072 \\
        Bitwise Operators  & 2     &  96   &  2    &  96   \\
        Registers          & 108   &  5184 &  267  &  12816   \\
        \hline
    \end{tabular}}
\end{table}

\subsection{Performance of General-Purpose Cores}
\begin{figure}[h!]
    \centering
    \includegraphics[width=0.92\linewidth]{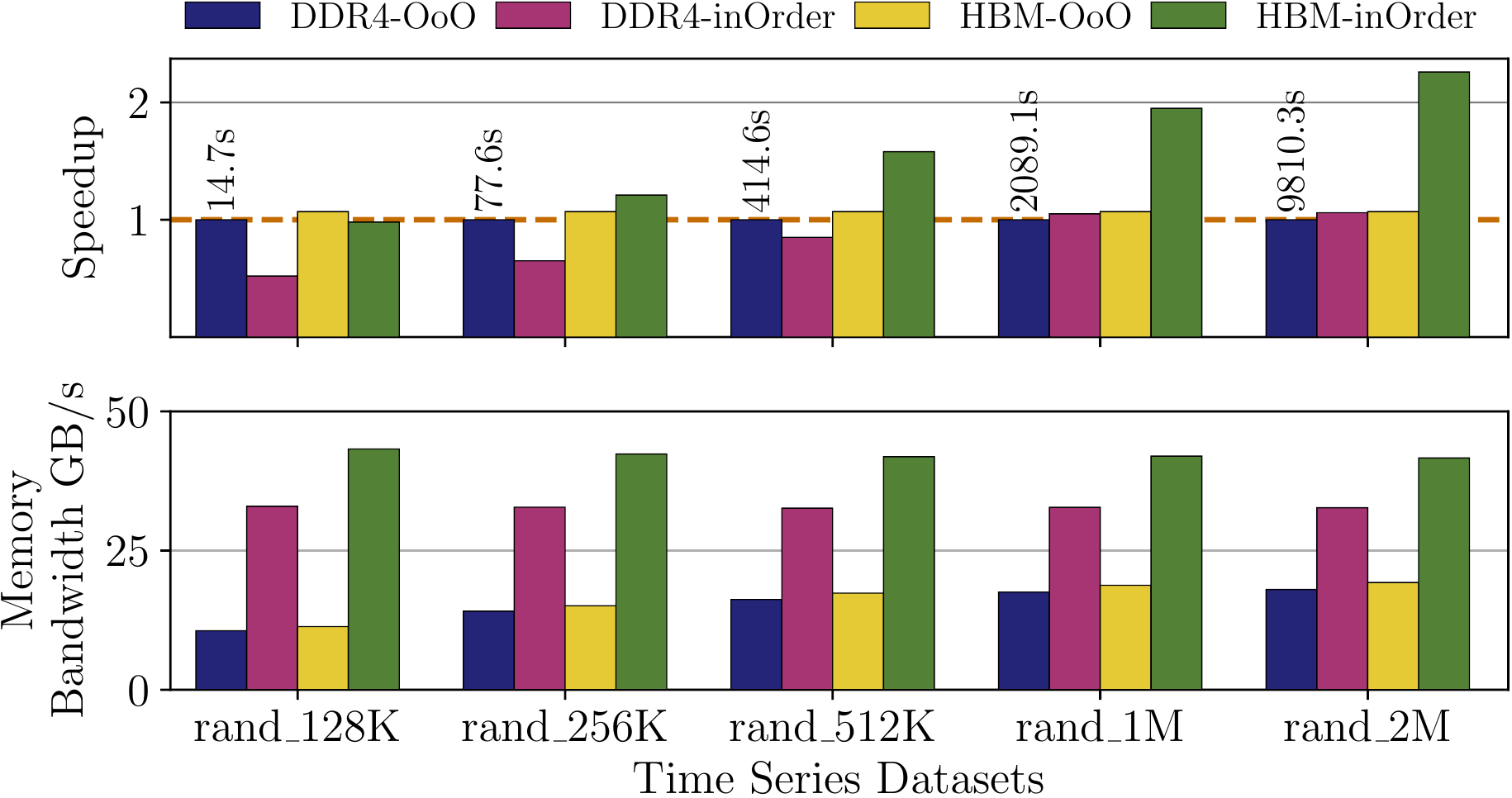}
    \caption{Speedup over the baseline DDR4-OoO and memory bandwidth usage for general-purpose platforms.}
    \vspace{-1mm}
    \label{fig:speedup}
\end{figure}

We evaluate the speedup over the baseline (DDR4-OoO) and memory bandwidth usage of SCRIMP, calculated using the ZSim-Ramulator framework for the \mbox{DDR4-OoO}, DDR4-inOrder, HBM-OoO and HBM-inOrder platforms using double-precision time series of different lengths ($n$), in Fig.~\ref{fig:speedup}. 
 
We report execution time of the baseline (DDR4-OoO) on top of the respective performance bars in Fig.~\ref{fig:speedup}. Based on these results, we make three key observations. 
First, the DDR4-OoO platform does not use the peak available bandwidth of DDR4 (i.e., 38.4GB/s). We reinforce this observation with our HBM-OoO evaluation which replaces DDR4 with higher bandwidth HBM. HBM-OoO platform improves performance by only 7\%, which means that providing more bandwidth does not significantly affect performance. This is because both platforms are compute-bound when executing SCRIMP. Second, the 64 lightweight cores of DDR4-inOrder slightly outperform the 8 complex cores of DDR4-OoO when $n\geq1048576$ elements (i.e., rand\_1M dataset). This is because shorter time series can fit in the L3 cache. For long time series, the higher parallelism provided by the in-order platform enables higher memory-level parallelism~\cite{mutlu2003runahead,mutlu2005techniques, Glew1998MLPYI, mutlu2008parallelism,1261383,mutlu2003micro} and higher memory bandwidth demand, where DDR4 bandwidth becomes a bottleneck, resulting in a memory-bound system. Third, the HBM-inOrder platform provides up to 2.25$\times$ speedup over the baseline (DDR4-OoO), and consumes only 17\% of the HBM's peak bandwidth with the largest dataset evaluated. In this case, even though performance is improved, the application is still compute-bound and simple NDP general-purpose cores cannot fully exploit the bandwidth provided by HBM (256GB/s)\footnote{Based on the memory bandwidth usage and McPAT, we estimate that a general-purpose based architecture would need 128 OoO cores (area 688mm\textsuperscript{2}, TDP 1137W, 18nm) or 384 in-order cores (area 164mm\textsuperscript{2}, TDP 126W, 18nm) to take full advantage of the maximum bandwidth provided by HBM.} for the largest dataset we evaluate, which means that large datasets can be comfortably accommodated. 

We conclude that general-purpose platforms provide less performance than NATSA's balanced design because they do not effectively exploit the memory bandwidth of HBM.

\subsection{Accuracy and Sensitivity to Window Size}
\label{sec:accuracy_window}

\begin{figure}[h]
    \centering
    \includegraphics[width=0.95\linewidth]{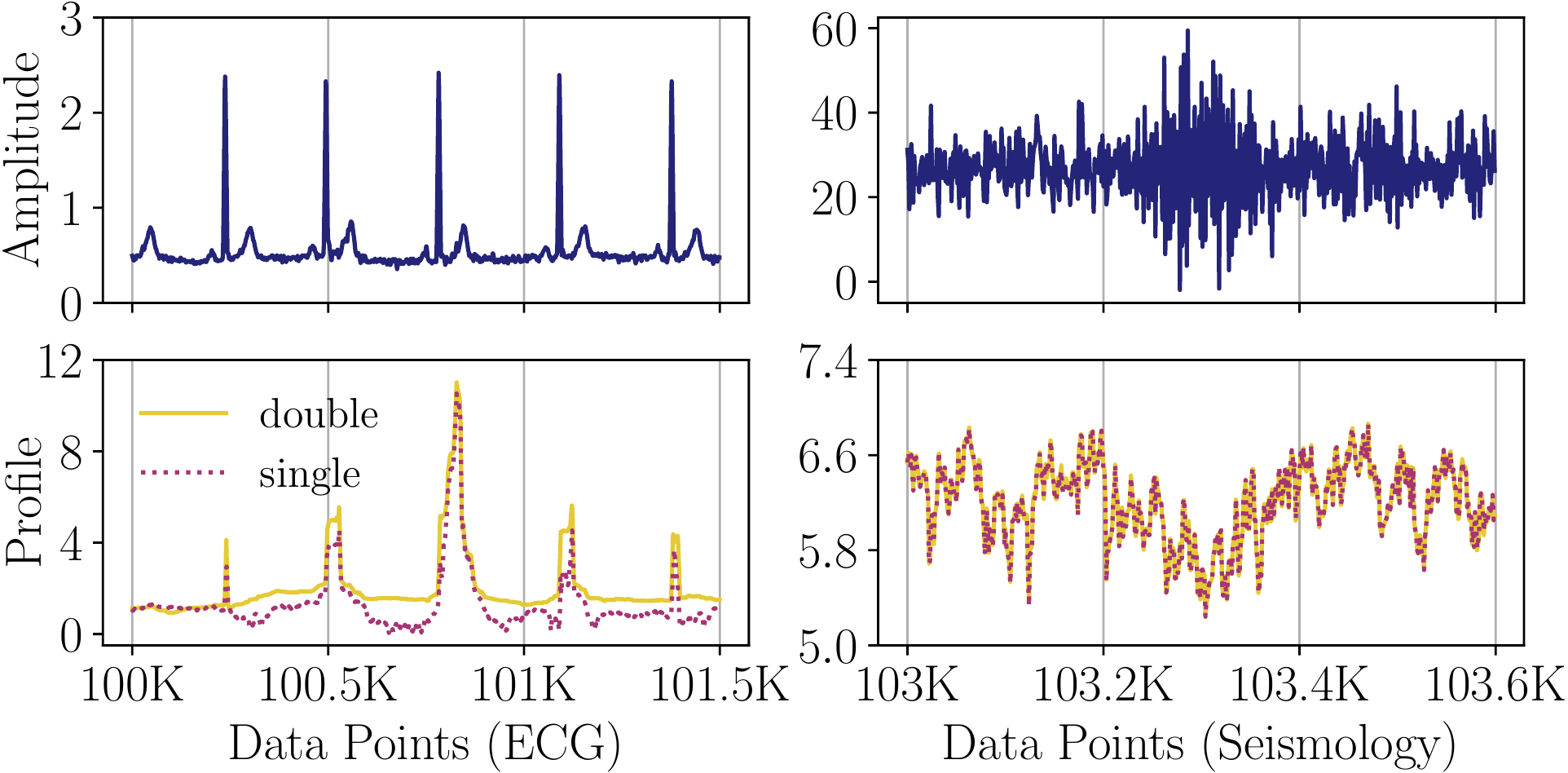}
    \caption{ECG (left) and seismology (right) data along with their profiles, calculated by NATSA using double and single precision, where events are easily visible.}
    \label{fig:precision_example}
\end{figure}

\textbf{Accuracy}. We explore how the accuracy of the SCRIMP implementation is affected by changing the precision of the data representation. We use real data obtained from~\cite{TDE+92} and~\cite{YHK16}, as discussed in Section~\ref{sec:mprofLoad}. 
Fig.~\ref{fig:precision_example} presents the output obtained for an electrocardiogram (ECG) and for seismology data using two precision values. We observe that events are still detectable even when reducing the precision from \textit{double} to \textit{single} precision. This observation can be exploited to improve performance and reduce energy consumption, by operating on smaller arithmetic units and less memory footprint. 

\noindent \textbf{Sensitivity to the subsequence length}.
We also perform a sensitivity analysis to the subsequence length ($m$). We observe that, when the proportion between $m$ and $n$ is less than two orders of magnitude, the performance of SCRIMP in all platforms is significantly affected by $m$. For example, when increasing $m$ from 1,024 to 16,384 in a time series of 131,072 elements, the execution time of SCRIMP reduces by 41\%. However, when the time series length is large enough compared to the subsequence length, performance of SCRIMP is affected by a smaller amount. For instance, when increasing $m$ from 1,024 to 16,384 in a time series of 2,097,152 elements, the execution time of SCRIMP reduces by 13\%. This is because the computation of the first element of each diagonal involves the dot product calculation without any reutilization.

\section{Related Work}
To our knowledge, this is the first work that proposes a near-data processing accelerator for time series analysis. In this section, we briefly discuss prior work related to time series motif discovery and application-specific NDP accelerators.

Multiple techniques exist for time  series motif and discord discovery~\cite{patel2002mining, CKL03, tanaka2005discovery, ferreira2006mining, yankov2007detecting, tang2008discovering, MKZ+09,castro2010multiresolution, nunthanid2011discovery, nunthanid2012parameter, li2012visualizing, mueen2015enumeration, yingchareonthawornchai2013efficient, gulati2014mining, torkamani2015shift, balasubramanian2016discovering}. A survey on time series motif discovery algorithms can be found in~\cite{TL17}. These implementations are approximate or exact~\cite{M14} in finding motifs and discords, which affects the time complexity of the algorithm. Exact motif and discord discovery processing of exceptionally large time series can be very time-consuming~\cite{MPROFILEXIV}. Consequently, \emph{anytime} algorithms~\cite{MPROFILEI} are proposed to return a valid solution even if they are interrupted, and are expected to find better solutions the longer they run. \emph{Matrix profile}~\cite{MPROFILEI} is the state-of-the-art exact \emph{anytime} algorithm for time series motif and discord discovery. There are several implementations of \emph{matrix profile}, including STAMP~\cite{MPROFILEI}, STOMP~\cite{MPROFILEII}, SCRIMP~\cite{MPROFILEXI} and SCAMP~\cite{MPROFILEXIV}. SCRIMP is the state-of-the-art CPU-based implementation. Prior acceleration approaches to time series analysis~\cite{MPROFILEII,MPROFILEXI} mainly focus on accelerating STOMP and PreSCRIMP~\cite{MPROFILEXI} on GPUs. Recently, SCAMP~\cite{MPROFILEXIV} framework combines a host (either a local machine or a server in a compute cluster) and workers that follow the directions from the host (either other CPUs in the cluster or accelerators such as GPUs). A SCRIMP version tuned for a many-core CPU (Intel Xeon Phi KNL) using vectorization can be found in~\cite{FVG+19}.

Recent works explore Near Data Processing~\cite{mutlu2019processing, ghose2018enabling, ghose2019processing,Hsieh2016accelerating,wong2012metal,aga2017compute,asghari2016chameleon,chi2016prime,hashemi2016continuous,KKC+16,loh2013processing,seshadri2019dram,ahn2015pim,seshadri2017simple, GAK15, DDM+17, AHY+15, GPY+17, LMd+18, Boroumand2018google, boroumand2019conda, boroumand2017lazypim, Hsieh2016TOM,Kim2018Grim,cali2020genasm,hashemi2016accelerating,seshadri2015fast,li2016pinatubo,singh2019napel,pattnaik2016scheduling,seshadri2017ambit,singh2020nero,hashemi2016accelerating,seshadri2013rowclone,qiao2018atomlayer,chang2016low} for various applications using accelerators or general-purpose cores. In~\cite{DDM+17}, ARM cores are used as NDP compute units to improve data analytics operators (e.g., group, join, sort). IMPICA~\cite{Hsieh2016accelerating} is an NDP pointer chasing accelerator. Tesseract~\cite{AHY+15} is a scalable NDP accelerator for parallel graph processing. TETRIS~\cite{GPY+17} is an  NDP neural network accelerator. Lee et al.~\cite{LMd+18} propose an NDP accelerator for similarity search. GRIM-Filter~\cite{Kim2018Grim} is an NDP accelerator for pre-alignment filtering~\cite{xin2012fasthash,alser2019shouji,xin2015shifted,alser2019sneakysnake,Alser2017GateKeeper} in genome analysis~\cite{alser2020accelerating}. Boroumand et al.~\cite{Boroumand2018google} analyze the energy and performance impact of data movement for several widely-used Google consumer workloads, providing NDP accelerators for them. CoNDA~\cite{boroumand2019conda} provides efficient cache coherence support for NDP accelerators. Finally, an NDP architecture~\cite{PJZ+14} has been proposed for MapReduce-style applications.

\section{Conclusion}
We introduce NATSA, the first Near-Data-Processing (NDP) accelerator for time series analysis. NATSA 1) exploits the memory bandwidth of high-bandwidth memory (HBM) to analyze time series data at scale for a wide range of applications, 2) improves energy efficiency
and execution time by using specialized low-power arithmetic units close to  HBM memory, and 3) provides a novel workload scheduling scheme to prevent load imbalance and preserve the \emph{anytime} property. NATSA outperforms the hardware platforms we evaluate in terms of performance, energy consumption and area requirements.
We conclude that NATSA is an efficient NDP accelerator for time series, and hope that this work inspires future research directions in NDP for time series analysis.

\section*{Acknowledgments}
We thank the anonymous reviewers of ICCD 2020 for feedback. This work has been supported by TIN2016-80920-R and UMA18-FEDERJA-197 Spanish projects, and Eurolab4HPC and HiPEAC collaboration grants. We also acknowledge support from the SAFARI Group's industrial partners, especially ASML, Facebook, Google, Huawei, Intel, Microsoft, and VMware, as well as support from Semiconductor Research Corporation.


\bibliographystyle{IEEEtranS}
\bibliography{mybibfile}

\begin{thebibliography}{100}
\providecommand{\url}[1]{#1}
\csname url@samestyle\endcsname
\providecommand{\newblock}{\relax}
\providecommand{\bibinfo}[2]{#2}
\providecommand{\BIBentrySTDinterwordspacing}{\spaceskip=0pt\relax}
\providecommand{\BIBentryALTinterwordstretchfactor}{4}
\providecommand{\BIBentryALTinterwordspacing}{\spaceskip=\fontdimen2\font plus
\BIBentryALTinterwordstretchfactor\fontdimen3\font minus
  \fontdimen4\font\relax}
\providecommand{\BIBforeignlanguage}[2]{{%
\expandafter\ifx\csname l@#1\endcsname\relax
\typeout{** WARNING: IEEEtranS.bst: No hyphenation pattern has been}%
\typeout{** loaded for the language `#1'. Using the pattern for}%
\typeout{** the default language instead.}%
\else
\language=\csname l@#1\endcsname
\fi
#2}}
\providecommand{\BIBdecl}{\relax}
\BIBdecl

\bibitem{PCM}
``{Intel Processor Counter Monitor},'' \url{https://github.com/opcm/pcm},
  accessed 23 September 2020.

\bibitem{MICRONPOWERCALCS}
``Micron {P}ower {C}alculator,''
  \url{www.micron.com/support/tools-and-utilities/power-calc}, accessed 23
  September 2020.

\bibitem{GTX1050}
``{NVIDIA GTX 1050 Specs},''
  \url{https://www.nvidia.com/en-in/geforce/products/10series/geforce-gtx-1050/},
  accessed 23 September 2020.

\bibitem{NVVP}
``{NVIDIA} {V}isual {P}rofiler,''
  \url{https://developer.nvidia.com/nvidia-visual-profiler}, accessed 23
  September 2020.

\bibitem{aga2017compute}
S.~Aga \emph{et~al.}, ``Compute caches,'' in \emph{HPCA}, 2017.

\bibitem{AHY+15}
J.~Ahn \emph{et~al.}, ``{A Scalable Processing-In-Memory Accelerator for
  Parallel Graph Processing},'' in \emph{ISCA}, 2015.

\bibitem{ahn2015pim}
J.~Ahn \emph{et~al.}, ``{{PIM}-Enabled Instructions: A Low-Overhead,
  Locality-Aware Processing-In-Memory Architecture},'' in \emph{ISCA}, 2015.

\bibitem{alser2020accelerating}
M.~Alser \emph{et~al.}, ``{Accelerating Genome Analysis: A Primer on an Ongoing
  Journey},'' \emph{IEEE Micro}, 2020.

\bibitem{alser2019shouji}
M.~Alser \emph{et~al.}, ``{Shouji: A Fast and Efficient Pre-Alignment Filter
  for Sequence Alignment},'' \emph{Bioinformatics}, 2019.

\bibitem{Alser2017GateKeeper}
M.~Alser \emph{et~al.}, ``{{GateKeeper: A New Hardware Arch. for Accelerating
  Pre-alignment in DNA Short Read Mapping}},'' \emph{Bioinformatics}, 2017.

\bibitem{alser2019sneakysnake}
M.~Alser \emph{et~al.}, ``{SneakySnake: A Fast and Accurate Universal Genome
  Pre-Alignment Filter for CPUs, GPUs, and FPGAs},'' \emph{arXiv}, 2019.

\bibitem{asghari2016chameleon}
H.~Asghari-Moghaddam \emph{et~al.}, ``{Chameleon: Versatile and Practical
  Near-DRAM Acceleration Arch. for Large Mem. Sys.}'' in \emph{MICRO}, 2016.

\bibitem{balasubramanian2016discovering}
A.~Balasubramanian \emph{et~al.}, ``{Discovering Multidimensional Motifs in
  Physiological Signals for Personalized Healthcare},'' \emph{JSTSP}, 2016.

\bibitem{B04}
Z.~Bar-Joseph, ``{Analyzing Time Series Gene Expression Data},''
  \emph{Bioinformatics}, 2004.

\bibitem{BBB+11}
N.~Binkert \emph{et~al.}, ``{The gem5 Simulator},'' \emph{Comp. Arch. News},
  2011.

\bibitem{Boroumand2018google}
A.~Boroumand \emph{et~al.}, ``{Google Workloads for Consumer Devices:
  Mitigating Data Movement Bottlenecks},'' \emph{ASPLOS}, 2018.

\bibitem{boroumand2019conda}
A.~Boroumand \emph{et~al.}, ``{CoNDA: Efficient Cache Coherence Support for
  Near-Data Accelerators},'' in \emph{ISCA}, 2019.

\bibitem{boroumand2017lazypim}
A.~Boroumand \emph{et~al.}, ``{LazyPIM: Efficient Support for Cache Coherence
  in Processing-In-Memory Architectures},'' \emph{arXiv}, 2017.

\bibitem{cali2020genasm}
D.~S. Cali \emph{et~al.}, ``{GenASM: A High-Performance, Low-Power Approximate
  String Matching Acceleration Framework for Genome Sequence Analysis},'' in
  \emph{MICRO}, 2020.

\bibitem{CCR19}
E.~Cartwright \emph{et~al.}, ``{Financial Time Series: Motif Discovery and
  Analysis Using VALMOD},'' in \emph{ICCS}, 2019.

\bibitem{CAC+13}
C.~Cassisi \emph{et~al.}, ``{Motif Discovery on Seismic Amplitude T. Series:
  The Case Study of Mt Etna 2011 Eruptive Activity},'' \emph{Pure Appl.
  Geophy.}, 2013.

\bibitem{castro2010multiresolution}
N.~Castro \emph{et~al.}, ``{Multireso. Motif Disco. in Time Series},'' in
  \emph{SDM}, 2010.

\bibitem{chang2016low}
K.~K. Chang \emph{et~al.}, ``{Low-Cost Inter-Linked Subarrays (LISA): Enabling
  Fast Inter-Subarray Data Movement in DRAM},'' in \emph{HPCA}, 2016.

\bibitem{chi2016prime}
P.~Chi \emph{et~al.}, ``{PRIME: A Novel Processing-In-Memory Arch. for Neural
  Network Computation in ReRAM-Based Main Memory},'' in \emph{ISCA}, 2016.

\bibitem{CKL03}
B.~Chiu \emph{et~al.}, ``{Probabilistic Discovery of Time Series Motifs},'' in
  \emph{SIGKDD}, 2003.

\bibitem{DDM+17}
M.~P. Drumond \emph{et~al.}, ``{The Mondrian Data Engine},'' in \emph{ISCA},
  2017.

\bibitem{FVG+19}
I.~Fernandez \emph{et~al.}, ``{Accelerating Time Series Motif Discovery in the
  {Intel Xeon Phi KNL} Processor},'' \emph{The Journal of Supercomputing},
  2019.

\bibitem{ferreira2006mining}
P.~G. Ferreira \emph{et~al.}, ``{Mining Approximate Motifs in Time Series},''
  in \emph{International Conference on Discovery Science}, 2006.

\bibitem{GH10}
S.~Galal \emph{et~al.}, ``{Energy-Efficient Floating-Point Unit Design},''
  \emph{IEEE Transactions on Computers}, 2010.

\bibitem{GAK15}
M.~Gao \emph{et~al.}, ``{Practical Near-Data Processing for In-Memory Analytics
  Frameworks},'' in \emph{PACT}, 2015.

\bibitem{GPY+17}
M.~Gao \emph{et~al.}, ``{{TETRIS:} Scalable and Efficient Neural Network
  Acceleration with 3{D} Memory},'' in \emph{ASPLOS}, 2017.

\bibitem{GNN+17}
P.~Garrard \emph{et~al.}, ``{Motif Discovery in Speech: Application to
  Monitoring Alzheimer’s Disease},'' \emph{Current Alzheimer Research}, 2017.

\bibitem{ghose2019processing}
S.~Ghose \emph{et~al.}, ``{Processing-In-Memory: A Workload-Driven
  Perspective},'' \emph{IBM Journal of Research and Development}, 2019.

\bibitem{ghose2018enabling}
S.~Ghose \emph{et~al.}, ``{Enabling the Adoption of Processing-In-Memory:
  Challenges, Mechanisms, Future Research Directions},'' \emph{arXiv}, 2018.

\bibitem{Ghose2019demystifying}
S.~Ghose \emph{et~al.}, ``{Demystifying Complex Workload-DRAM Interactions: An
  Experimental Study},'' in \emph{SIGMETRICS}, 2019.

\bibitem{Glew1998MLPYI}
A.~Glew, ``{MLP yes! ILP no!}'' in \emph{ASPLOS}, 1998.

\bibitem{gulati2014mining}
S.~Gulati \emph{et~al.}, ``{Mining Melodic Patterns in Large Audio Collections
  of Indian Art Music},'' in \emph{SITIS}, 2014.

\bibitem{PJZ+14}
S.~H~Pugsley \emph{et~al.}, ``{{NDC}: Analyzing the Impact of 3{D}-stacked
  Memory+Logic Devices on {M}ap{R}educe Workloads},'' in \emph{ISPASS}, 2014.

\bibitem{hadidi2017demystifying}
R.~Hadidi \emph{et~al.}, ``{Demystifying the Characteristics of 3D-stacked
  Memories: A case Study for Hybrid Memory Cube},'' in \emph{IISWC}, 2017.

\bibitem{hashemi2016accelerating}
M.~Hashemi \emph{et~al.}, ``{Accelerating Dependent Cache Misses with an
  Enhanced Memory Controller},'' in \emph{ISCA}, 2016.

\bibitem{hashemi2016continuous}
M.~Hashemi \emph{et~al.}, ``{Continuous Runahead: Transparent Hardware
  Acceleration for Memory Intensive Workloads},'' in \emph{MICRO}, 2016.

\bibitem{Hsieh2016TOM}
K.~Hsieh \emph{et~al.}, ``{ TOM: Enabling Programmer-Transparent Near-Data
  Processing in GPU Systems},'' in \emph{ISCA}, 2016.

\bibitem{Hsieh2016accelerating}
K.~Hsieh \emph{et~al.}, ``{Accelerating Pointer Chasing in 3D-stacked Memory:
  Challenges, Mechanisms, Evaluation},'' in \emph{ICCD}, 2016.

\bibitem{MPROFILEII}
Y.~Hu \emph{et~al.}, ``{Matrix {P}rofile {II}: Exploiting a Novel Algorithm and
  {GPUs} to Break the One Hundred Million Barrier for Time Series Motifs and
  Joins},'' in \emph{ICDM}, 2016.

\bibitem{HAA+17}
L.~Hussain \emph{et~al.}, ``{Symbolic Time Series Analysis of ({EEG}) Epileptic
  Seizure and Brain Dynamics with Eye-Open and Eye-Closed Subjects During
  Resting States},'' \emph{Journal of Physiological Anthropology}, 2017.

\bibitem{DDR4}
{JEDEC JESD79-4C}, ``{DDR4} {SDRAM} standard,''
  \url{www.jedec.org/standards-documents/docs/jesd79-4a}, accessed 23 September
  2020.

\bibitem{JCL+17}
H.~Jun \emph{et~al.}, ``{{HBM DRAM} Technology and Architecture},'' in
  \emph{IMW}, 2017.

\bibitem{KLL+06}
E.~Keogh \emph{et~al.}, ``{Finding the Most Unusual Time Series Subsequence:
  Algorithms and Applications},'' \emph{Knowledge and Information Systems},
  2006.

\bibitem{KKC+16}
D.~Kim \emph{et~al.}, ``{Neurocube: A Programmable Digital Neuromorphic
  Architecture with High-density 3D Memory},'' in \emph{ISCA}, 2016.

\bibitem{kim2018dram}
J.~S. Kim \emph{et~al.}, ``{The DRAM latency PUF: Quickly Evaluating Physical
  Unclonable Functions by Exploiting the Latency-Reliability Tradeoff in Modern
  Commodity DRAM Devices},'' in \emph{HPCA}, 2018.

\bibitem{kim2019d}
J.~S. Kim \emph{et~al.}, ``{D-RaNGe: Using Com. DRAM Devices to Generate True
  Random Numb. with Low Lat. and High Throughput},'' in \emph{HPCA}, 2019.

\bibitem{Kim2018Grim}
J.~S. Kim \emph{et~al.}, ``{GRIM-Filter: Fast seed Location Filter. in DNA Read
  Mapping Using PIM Technologies},'' \emph{BMC Genomics}, 2018.

\bibitem{KYM15}
Y.~Kim \emph{et~al.}, ``{Ramulator: A Fast and Extensible {DRAM} Simulator},''
  \emph{CAL}, 2015.

\bibitem{LCD04}
A.~Lakhina \emph{et~al.}, ``{Characterization of Network-Wide Anomalies in
  Traffic Flows},'' in \emph{IMC}, 2004.

\bibitem{HBM}
D.~U. {Lee} \emph{et~al.}, ``{25.2 A 1.2{V} 8{GB} 8-channel 128{GB}/s
  High-Bandwidth Memory ({HBM}) Stacked {DRAM} with Effective Microbump {I/O}
  Test Methods using 29nm Process and {TSV}},'' in \emph{ISSCC}, 2014.

\bibitem{Lee2016Simultaneous}
D.~Lee \emph{et~al.}, ``{Simultaneous Multi-Layer Access: Improving 3D-Stacked
  Memory Bandwidth at Low Cost},'' \emph{TACO}, 2016.

\bibitem{LMd+18}
V.~T. Lee \emph{et~al.}, ``{Application Codesign of NDP for Similarity
  Search},'' in \emph{IPDPS}, 2018.

\bibitem{LWT+19}
K.~H.~C. Li \emph{et~al.}, ``{The Current State of Mobile Phone Apps for
  Monitoring Heart Rate, Heart Rate Variability, and Atrial Fibrillation:
  Narrative Review},'' \emph{JMIR Mhealth Uhealth}, 2019.

\bibitem{LAS+09}
S.~Li \emph{et~al.}, ``{{McPAT}: An Integrated Power, Area, and Timing Modeling
  Framework for Multicore and Manycore Architectures},'' in \emph{MICRO}, 2009.

\bibitem{li2016pinatubo}
S.~Li \emph{et~al.}, ``{Pinatubo: A Processing-in-Memory Architecture for Bulk
  Bitwise Operations in Emerging Non-volatile Memories},'' in \emph{DAC}, 2016.

\bibitem{li2012visualizing}
Y.~Li \emph{et~al.}, ``{Visualizing Variable-Length Time Series Motifs},'' in
  \emph{SDM}, 2012.

\bibitem{loh2013processing}
G.~H. Loh \emph{et~al.}, ``{A Processing in Memory Taxonomy and a Case for
  Studying Fixed-Function PIM},'' in \emph{WoNDP}, 2013.

\bibitem{LCM+05}
C.-K. Luk \emph{et~al.}, ``{Pin: Building Customized Program Analysis Tools
  with Dynamic Instrumentation},'' in \emph{PLDI}, 2005.

\bibitem{MRB+11}
A.~McGovern \emph{et~al.}, ``{Identifying Predictive Multi-Dimensional Time
  Series Motifs: An Application to Severe Weather Prediction},'' \emph{Data
  Mining and Knowledge Discovery}, 2011.

\bibitem{M14}
A.~Mueen, ``{Time Series Motif Discovery: Dimensions and Applications},''
  \emph{WIREs: Data Mining and Knowledge Discovery}, 2014.

\bibitem{mueen2015enumeration}
A.~Mueen \emph{et~al.}, ``{Enumeration of Time Series Motifs of All Lengths},''
  \emph{Knowledge and Information Systems}, 2015.

\bibitem{MKZ+09}
A.~Mueen \emph{et~al.}, ``{Exact Discovery of Time Series Motifs},'' in
  \emph{SDM}, 2009.

\bibitem{mutlu2019processing}
O.~Mutlu \emph{et~al.}, ``{Processing Data Where it Makes Sense: Enabling
  In-Memory Computation},'' \emph{Microprocessors and Microsystems}, 2019.

\bibitem{mutlu2005techniques}
O.~Mutlu \emph{et~al.}, ``{Techniques for Efficient Processing in Runahead
  Execution Engines},'' in \emph{ISCA}, 2005.

\bibitem{1261383}
O.~Mutlu \emph{et~al.}, ``{Efficient Runahead Execution: Power-Efficient Memory
  Latency Tolerance},'' \emph{IEEE Micro}, 2006.

\bibitem{mutlu2008parallelism}
O.~Mutlu \emph{et~al.}, ``{Parallelism-Aware Batch Scheduling: Enhancing Both
  Performance and Fairness of Shared {DRAM} Systems},'' in \emph{ISCA}, 2008.

\bibitem{mutlu2003runahead}
O.~Mutlu \emph{et~al.}, ``{Runahead Execution: An Alternative to Very Large
  Instruction Windows for Out-of-Order Processors},'' in \emph{HPCA}, 2003.

\bibitem{mutlu2003micro}
O.~Mutlu \emph{et~al.}, ``{Runahead Execution: An Effective Alternative to
  Large Instruction Windows},'' \emph{IEEE Micro}, 2003.

\bibitem{nunthanid2011discovery}
P.~Nunthanid \emph{et~al.}, ``{Discovery of Variable Length Time Series
  Motif},'' in \emph{ECTI-CON}, 2011.

\bibitem{nunthanid2012parameter}
P.~Nunthanid \emph{et~al.}, ``{Parameter-Free Motif Discovery for Time Series
  Data},'' in \emph{ECTI-CON}, 2012.

\bibitem{nvidia2013k40}
NVIDIA, ``{Tesla K40 GPU Active Accelerator},'' \emph{Board specification},
  2013.

\bibitem{patel2002mining}
P.~Patel \emph{et~al.}, ``{Mining Motifs in Massive Time Series Databases},''
  in \emph{ICDM}, 2002.

\bibitem{pattnaik2016scheduling}
A.~Pattnaik \emph{et~al.}, ``{Scheduling Techniques for GPU Architectures with
  Processing-in-Memory Capabilities},'' in \emph{PACT}, 2016.

\bibitem{qiao2018atomlayer}
X.~Qiao \emph{et~al.}, ``{Atomlayer: a Universal Reram-based CNN Accelerator
  with Atomic Layer Computation},'' in \emph{DAC}, 2018.

\bibitem{RGS+15}
G.~Radhakrishnan \emph{et~al.}, ``{Experimentation and Analysis of Time Series
  Data from Multi-Path Robotic Environment},'' in \emph{CONECCT}, 2015.

\bibitem{MILLIONTRILLIONS}
T.~Rakthanmanon \emph{et~al.}, ``{Searching and Mining Trillions of Time Series
  Subsequences Under Dynamic Time Warping},'' in \emph{KDD}, 2012.

\bibitem{RAMULATOR_GITHUB}
{{SAFARI Research Group}}, ``{Ramulator} {Source} {Code},''
  \url{https://github.com/CMU-SAFARI/ramulator}, accessed 23 September 2020.

\bibitem{salehi2015energy}
S.~Salehi \emph{et~al.}, ``{Energy and Area Analysis of a Floating-Point Unit
  in 15nm CMOS Process Technology},'' in \emph{SoutheastCon}, 2015.

\bibitem{SK13}
D.~Sanchez \emph{et~al.}, ``{{ZSim}: Fast and Accurate Microarchitectural
  Simulation of Thousand-Core Systems},'' in \emph{ISCA}, 2013.

\bibitem{SBP+12}
A.~Sathyanarayana \emph{et~al.}, ``{CAN-Bus Signal Analysis Using Stochastic
  Methods and Pattern Recognition in Time Series for Active Safety},''
  \emph{Springer-Verlag}, 2012.

\bibitem{seshadri2015fast}
V.~Seshadri \emph{et~al.}, ``{Fast Bulk Bitwise AND and OR in DRAM},''
  \emph{CAL}, 2015.

\bibitem{seshadri2013rowclone}
V.~Seshadri \emph{et~al.}, ``{RowClone: Fast and Energy-Efficient in-DRAM Bulk
  Data Copy and Initialization},'' in \emph{MICRO}, 2013.

\bibitem{seshadri2017ambit}
V.~Seshadri \emph{et~al.}, ``{Ambit: In-memory Accelerator for Bulk Bitwise
  Operations Using Commodity DRAM Technology},'' in \emph{MICRO}, 2017.

\bibitem{seshadri2017simple}
V.~Seshadri \emph{et~al.}, ``{Simple Operations in Memory to Reduce Data
  Movement},'' in \emph{Advances in Computers}.\hskip 1em plus 0.5em minus
  0.4em\relax Elsevier, 2017.

\bibitem{seshadri2019dram}
V.~Seshadri \emph{et~al.}, ``{In-{DRAM} Bulk Bitwise Execution Engine},''
  \emph{arXiv}, 2019.

\bibitem{SRW+14}
Y.~S. Shao \emph{et~al.}, ``{Aladdin: A {Pre-RTL}, Power-Performance
  Accelerator Simulator Enabling Large Design Space Exploration of Customized
  Architectures},'' in \emph{ISCA}, 2014.

\bibitem{SS17}
R.~H. Shumway \emph{et~al.}, ``{Time Series Analysis and Its Applications: With
  R Examples},'' 2017.

\bibitem{singh2020nero}
G.~Singh \emph{et~al.}, ``{NERO: A Near High-Bandwidth Memory Stencil
  Accelerator for Weather Prediction Modeling},'' in \emph{FPL}, 2020.

\bibitem{singh2019napel}
G.~Singh \emph{et~al.}, ``{NAPEL: Near-memory Computing Application Performance
  Prediction Via Ensemble Learning},'' in \emph{DAC}, 2019.

\bibitem{sodani2015knights}
A.~Sodani, ``{Knights Landing (KNL): 2nd Generation Intel{\textregistered} Xeon
  Phi Processor},'' in \emph{HCS}, 2015.

\bibitem{SXS+16}
S.~Y. Sophia \emph{et~al.}, ``{Co-Designing Accelerators and {SoC} Interfaces
  Using gem5-{Aladdin}},'' in \emph{MICRO}, 2016.

\bibitem{SDW15}
B.~Szigeti \emph{et~al.}, ``{Searching for Motifs in the Behaviour of Larval
  {Drosophila} Melanogaster and {Caenorhabditis Elegans} Reveals Continuity
  Between Behavioural States},'' \emph{Journal of The Royal Society}, 2015.

\bibitem{TDE+92}
A.~Taddei \emph{et~al.}, ``{{The European ST-T Database: Standard for
  Evaluating Systems for the Analysis of ST-T Changes in Ambulatory
  Electrocardiography}},'' \emph{European Heart Journal}, 1992.

\bibitem{tanaka2005discovery}
Y.~Tanaka \emph{et~al.}, ``{Discovery of Time-Series Motif from
  Multi-Dimensional Data Based on {MDL} Principle},'' \emph{Machine Learning},
  2005.

\bibitem{tang2008discovering}
H.~Tang \emph{et~al.}, ``{Discovering Original Motifs with Different Lengths
  from Time Series},'' \emph{Knowledge-Based Systems}, 2008.

\bibitem{TL17}
S.~Torkamani \emph{et~al.}, ``{Survey on Time Series Motif Discovery},''
  \emph{WIREs: Data Mining and Knowledge Discovery}, 2017.

\bibitem{torkamani2015shift}
S.~Torkamani \emph{et~al.}, ``{Shift-Invariant Feature Extraction for
  Time-Series Motif Discovery},'' in \emph{Workshop Computational
  Intelligence}, 2015.

\bibitem{wong2012metal}
H.-S.~P. Wong \emph{et~al.}, ``{Metal--oxide RRAM},'' \emph{Proceedings of the
  IEEE}, 2012.

\bibitem{xin2015shifted}
H.~Xin \emph{et~al.}, ``{Shifted Hamming Distance: A Fast and Accurate
  SIMD-friendly Filter to Accelerate Alignment Verification in Read Mapping},''
  \emph{Bioinformatics}, 2015.

\bibitem{xin2012fasthash}
H.~Xin \emph{et~al.}, ``{FastHASH: A New GPU-friendly Algorithm for Fast and
  Comprehensive Next-Generation Sequence Mapping},'' in \emph{BMC Genomics},
  2013.

\bibitem{yankov2007detecting}
D.~Yankov \emph{et~al.}, ``{Detecting Time Series Motifs Under Uniform
  Scaling},'' in \emph{SIGKDD}, 2007.

\bibitem{YHK16}
C.~M. Yeh \emph{et~al.}, ``{Matrix Profile III: The Matrix Profile Allows
  Visualization of Salient Subsequences in Massive Time Series},'' in
  \emph{ICDM}, 2016.

\bibitem{MPROFILEI}
C.-C.~M. Yeh \emph{et~al.}, ``{Matrix {P}rofile {I}: All Pairs Similarity Joins
  for Time Series: A Unifying View That Includes Motifs, Discords and
  Shapelets},'' in \emph{ICDM}, 2016.

\bibitem{YZU+18}
C.-C.~M. Yeh \emph{et~al.}, ``{Time Series Joins, Motifs, Discords and
  Shapelets: A Unifying View That Exploits The Matrix Profile},'' \emph{JDMKD},
  2018.

\bibitem{yingchareonthawornchai2013efficient}
S.~Yingchareonthawornchai \emph{et~al.}, ``{Efficient Proper Length Time Series
  Motif Discovery},'' in \emph{ICDM}, 2013.

\bibitem{zhang2014top}
D.~Zhang \emph{et~al.}, ``{{TOP-PIM}: Throughput-Oriented Programmable
  Processing in Memory},'' in \emph{HPDC}, 2014.

\bibitem{MPROFILEXI}
Y.~Zhu \emph{et~al.}, ``{Matrix {P}rofile {XI}: {SCRIMP}++: Time Series Motif
  Discovery at Interactive Speeds},'' in \emph{ICDM}, 2018.

\bibitem{MPROFILEXIV}
Z.~Zimmerman \emph{et~al.}, ``{Matrix {P}rofile {XIV}: Scaling Time Series
  Motif Discovery with {GPUs} to Break a Quintillion Pairwise Comparisons a Day
  and Beyond},'' in \emph{SoCC}, 2019.

\end{thebibliography}

\end{document}